\documentclass[11pt]{article}
\usepackage{epsfig}
\usepackage{amsmath}
\usepackage{amssymb}
\usepackage{amsfonts}
\usepackage{subfigure}
\usepackage{slashed}
\usepackage{xcolor}
\def\beq{\begin{equation}}
\def\eeq{\end{equation}}
\def\bea{\begin{eqnarray}}
\def\eea{\end{eqnarray}}
\def\ksl{\hbox{\hbox{${k}$}}\kern-1.9mm{\hbox{${/}$}}}

\newcommand{\nn}{\nonumber}

\def\lsim{\raise0.3ex\hbox{$\;<$\kern-0.75em\raise-1.1ex\hbox{$\sim\;$}}} 

\def\gsim{\raise0.3ex\hbox{$\;>$\kern-0.75em\raise-1.1ex\hbox{$\sim\;$}}}

\begin{document}

\begin{center}

{\bf \Large Fermion Scattering in a Gravitational Background: 

\vspace{0.2cm}
Electroweak Corrections and Flavour Transitions 
}
\vspace{0.3cm}

\vspace{1.5cm}
{\bf Claudio Corian\`{o}$^{a}$, Luigi Delle Rose$^{a}$, 
Emidio Gabrielli$^{b,c}$\footnote{
On leave of absence from Dipartimento di Fisica  Universit\`a di 
Trieste, Strada Costiera 11, I-34151 Trieste \\} and Luca Trentadue$^{d}$}
\vspace{1cm}

{\it $^{(a)}$Dipartimento di Matematica e Fisica "Ennio De Giorgi", 
Universit\`{a} del Salento and \\ INFN-Lecce, Via Arnesano, 73100 Lecce, Italy\footnote{claudio.coriano@le.infn.it, luigi.dellerose@le.infn.it, 
emidio.gabrielli@cern.ch, luca.trentadue@cern.ch}}\\

\vspace{1cm}
{\it $^{(b)}$ NICPB, R\"avala 10, Tallinn 10143, Estonia\\
$^{(c)}$ INFN, Sezione di Trieste, Via Valerio 2, I-34127 Trieste, Italy \\}
\vspace{1cm}
{\it$^{(d)}$ Dipartimento di Fisica e Scienze della Terra "Macedonio Melloni", Universit\`a di Parma and
INFN, Sezione di Milano Bicocca, Milano, Italy
\\
}
\vspace{.5cm}
\begin{abstract} 

We investigate the role of the electroweak corrections to the scattering cross section of Standard Model fermions 
with gravity. We use both an approach of scattering off an external potential, where the gravitational field is treated as a classical background generated by a heavy source, and the usual interaction based on the one-graviton-exchange. In the potential appoach we consider the fields both of a localized and of a distributed gravitational source of spherical symmetry 
and uniform density, separating the cases of interactions taking place both in the inner and external regions of the source. This allows to make a distinction between interactions involving neutrinos and dark matter particles with a realistic gravity source, which cover the inner region, and the rest of the Standard model fermions. The role of the gravitationally induced flavour-changing transitions, as well as the flavour diagonal ones, are investigated in the limit of both large and small momentum transfers, deriving the structure of the corresponding Hamiltonian.  
\color{black}
\end{abstract}
\end{center}

\newpage

\section{Introduction}
Radiative corrections to the graviton/matter vertex have been investigated since the 70' s  
by Berends and Gastmans \cite{Berends:1975ah} in the QED case, who quantified the impact of these corrections to the bending of photons and fermions in an external localized gravitational field. These results were based on the study of 3-point functions (gravitational form factors) with two fermions ($Tff$) or two photons ($TVV$) in external gravity, and one insertion of the energy momentum tensor (EMT)  $(T)$ of the matter fields.  
 
 The very small size of these corrections, which affect the angles of deflection, combined with the experimental difficulty to improve on their direct measurements, call into question the possibility of using these effects as a possible test of General Relativity. Perhaps, it is for this reason that the study of the electroweak corrections to the bending angles have not drawn any attention on these processes until the analysis of \cite{Degrassi:2008mw} and, more  recently, of \cite{Coriano:2012cr, Coriano:2013msa}, which concern the fermion/graviton vertex.
 
 In the photon case, the $TVV$ vertex in QED has been re-investigated in the analysis of the conformal anomaly action in \cite{Giannotti:2008cv, Armillis:2009pq}, later extended to the electroweak case in \cite{Coriano:2011ti, Coriano:2011zk}. 
 In the $Tf f$ case, the study has been separated into the flavour-changing \cite{Degrassi:2008mw, Coriano:2013msa} and flavour diagonal sectors \cite{Coriano:2012cr}, the first of the two playing a significant role in the  analysis of possible flavour transitions in the presence of a scalar component in the external gravitational field, as pointed out in \cite{Degrassi:2008mw}. This aspect will be re-addressed in our current study. 
 
 The issue concerning the size of these corrections and their relevance can be resolved by noticing that the very same interactions become significant in the presence of sizeable gravitational backgrounds. Such are those due 
to black holes, which are ubiquitous in galactic centers, and surging to remarkable relevance in the analysis of galaxy structure formation.  
Obviously  our formalism, which simply extends to the electroweak case the original analysis of \cite{Berends:1975ah}, remains valid only in the case of small fluctuations around a flat spacetime metric, where curvature effects are ignored, and does not allow to handle backgrounds characterized by a strong curvature. It could, however,  be extended to include further corrections related to the curvature scale of certain backgrounds, in the limit in which gravity is treated as an external classical source. 
 
For this reason, we think that our analysis allows to highlight some of the basic features of the interaction of the Standard Model Lagrangian, once this is immersed in a weak gravitational background, clarifying some of the specific features of this coupling at a perturbative level. In particular, the renormalizability of the theory, which remains intact in this extended framework, is an important feature of these computations, not shared by other backgrounds. 

One of the specific features of our results is their possible application to the neutrino sector, which has received 
 some attention in recent years (see for instance \cite{Escribano:2001ew, Mena:2006ym, Eiroa:2008ks}).  We will provide indeed the expression for the QED and electroweak corrections to the scattering of a neutrino off an external source. 
These results are the starting point in the derivation of the equation relating the angle of deflection to the impact parameter of the collision, in the case of a static heavy source of the gravitational field. The use of an impact parameter formalism allows to merge the usual quantum corrections, computed in the exact electroweak theory, with the semiclassical method in which the incoming asymptotic quantum state is described by a particle approaching the source at a distance $b$, with $b$ denoting the impact parameter. This method has been used in \cite{Berends:1975ah} and in the more recent literature on the computation of corrections to the gravitational scattering of quantum fields both in static \cite{Accioly:2004bm, Sorge:2008zz} and in rotating (Lense-Thirring) backgrounds \cite{Sorge:2012cd}. Except for the case of the QED corrections, already discussed in \cite{Berends:1975ah} and that we recompute, a general analysis of the weak 
corrections and of the modified lense equations, which may impact the structure of cosmic shears, is left to future work.  
 
The primary goal of our work is to extend our previous analysis of the $Tff$ vertex, by computing the radiative corrections to the cross section in an external gravitational field, in the limit of a source of large mass, i.e. in a scattering off a potential. In this case we discuss both the limits of low and high momentum transfers. At low momentum transfers we derive an effective Hamiltonian which describes the flavour-changing transitions. We also discuss the cross section for the exchange of a scalar component for the gravitational field, with the inclusion of the electroweak effects. Our expressions for fermion scattering are specialized both to ordinary and to weakly interacting fermions, such as neutrinos or dark matter particles, by the inclusion of suitable form factors which account for interactions in the inner core of the source. We conclude our analysis with some perspectives and possible extensions of our work, to be left for further studies. Being this work the third in a sequel \cite{Coriano:2012cr, Coriano:2013msa}, most of our previous notations and results will be necessary in order to proceed with the applications discussed here.

\section{The Standard Model Lagrangian in a gravitational background: the fermion sector} 
In this section we briefly review the main features of the coupling of the Standard Model in an external gravitational field, focusing our attention on the fermion sector.
We recall that the dynamics of the Standard Model plus gravity is described by the action  
\beq
\mathcal{S} =  \mathcal{S}_{SM} +\mathcal{S}_G + \mathcal{S}_{I}
\eeq
with $\mathcal{S}_{SM}$ containing the Standard Model (SM) Lagrangian, while 
\bea
\mathcal{S}_G &=& -\frac{1}{\kappa^2}\int d^4 x \sqrt{-{g}}\, R \nn\\
\mathcal{S}_I &=&\chi \int d^4 x \sqrt{-{g}}\, R \, H^\dag H 
\label{thelagrangian}
\eea
denote respectively the Einstein gravitational term of the action and a term of improved action $\mathcal{S}_I$, involving the Higgs doublet $H$. The latter is responsible for generating a symmetric and traceless energy-momentum tensor (EMT)
\cite{Callan:1970ze}. $\mathcal{S}_{SM}$ is obtained by extending the ordinary Lagrangian of the Standard Model to a curved metric background. $\mathcal{S}_I$ vanishes in the flat space time limit, due to the vanishing of the Ricci scalar in the same limit. Its corresponding EMT $(T_{I\,\mu\nu}$), however, is non-vanishing. With the inclusion of $\mathcal{S}_I$, varying the action $\mathcal{S}$ with respect to the fields of the 
SM allows to generate an EMT which is symmetric without any additional symmetrization. We remark that $\chi$ is conformal at the value $\chi\equiv\chi_c=1/6$ and guarantees the renormalizability of the model, including all Green functions at any order in $k$ containing only insertions of external graviton fields $h$, with $\kappa^2=16\pi G$.

We will be using the flat metric $\eta_{\mu\nu}=\textrm{diag}(1,-1,-1,-1)$, with an expansion 
of the form
\bea
&& g_{\mu\nu}=\eta_{\mu\nu} + \kappa h_{\mu\nu} + O(\kappa^2)\nn \\
&& g^{\mu\nu}=\eta^{\mu\nu} - \kappa h^{\mu\nu}  +O(\kappa^2)\nn\\
&& \sqrt{-g}= 1+\frac{\kappa}{2}h + O(\kappa^2),
\eea
where $h\equiv h^{\mu\nu}\eta_{\mu\nu}$ is the trace of the metric fluctuations.

The interaction between the gravitational field and matter, at this order, is mediated by diagrams containing a single power of the EMT $T^{\mu\nu}$ and multiple fields of the Standard Model. The tree level coupling is summarized by the action 
\beq
\mathcal{S}_{int}=-\frac{\kappa}{2}\int d^4 x \, T_{\mu\nu} h^{\mu\nu} \,,
\label{inter}
\eeq
where $T_{\mu \nu}$ denotes the symmetric and covariantly conserved EMT of the Standard Model Lagrangian, embedded in a curved space-time background and defined as 
\beq
T_{\mu\nu}=\frac{2}{\sqrt{-g}}\frac{\delta \left(S_{SM}+S_{I}\right)}{\delta g^{\mu\nu}} \bigg|_{g=\eta} \,.
\eeq 

The complete EMT of the Standard Model, including ghost and gauge-fixing contributions can be found in \cite{Coriano:2011zk}. The fermionic part of the EMT is obtained using the vielbein formalism. Indeed the fermions are coupled to gravity by using the spin connection $\Omega$ induced by the curved metric $g_{\mu\nu}$.
This allows the introduction of a derivative $\mathcal D_{\mu}$ which is covariant both under gauge and diffeomorphism transformations.
Generically, the Lagrangian for a fermion $(f)$ takes the form
\bea
\mathcal L_{F} =\sqrt{-g} \left( \frac{i}{2} \bar \psi \gamma^\mu (\mathcal D_{\mu} \psi) - \frac{i}{2} (\mathcal D_{\mu} \bar \psi) \gamma^{\mu} \psi - m \, \bar \psi \psi \right) \,,
\label{dir}
\eea
where the covariant derivative is defined as $\mathcal D_{\mu} = \partial_{\mu} + A_{\mu} + \Omega_{\mu}$, with $A_{\mu}$ denoting the gauge field.
The spin connection takes the form
\bea
\Omega_{\mu} = \frac{1}{2} \sigma^{a b} V_a^{\nu} V_{b \nu ; \mu}
\eea
where $V$ is the vielbein, the semicolon denotes the gravitationally covariant derivative and $\sigma^{ab}$ are the generators of the Lorentz group in the spinorial representation. The latin indices are Lorentz indices of a local free-falling frame. The connection can be expanded as 
\beq
\Omega_{\mu}=\frac{1}{4}\sigma^{m n}\left[V^\nu_m\left(\partial_{\mu}V_{n\nu}-\partial_{\nu}V_{n\mu}\right)
 + \frac{1}{2}V^{\rho}_mV^{\sigma}_n\left( 
\partial_\sigma V_{l \rho}-\partial_{\rho}V_{l\sigma}\right) V_{\mu}^l-(m\leftrightarrow n)\right].
\eeq
with
\beq
V_\mu^m=\delta_\mu^m +\frac{\kappa}{2}h^m_\mu +O(\kappa^2).
\eeq
At leading order the interacting Dirac Lagrangian (\ref{dir}) generates the terms 
\bea
\mathcal{L}_F&=& \mathcal{L}_0-\frac{\kappa}{2}h_{\mu\nu}T^{(0) \mu\nu}
\eea
with $\mathcal{L}_0$ denoting the free Dirac term
\beq
\mathcal{L}_{0}=\frac{i}{2}\left(\bar{\psi}\stackrel{\rightarrow}{\slashed{\partial}}\psi-\bar{\psi}\stackrel{\leftarrow}{\slashed{\partial}}\psi  \right) - m \bar{\psi}\psi
\eeq
and the EMT of a Dirac fermion of mass $m_f$
\beq
T^{(0)}_{\mu\nu}=\frac{i}{4}\left((\bar{\psi}\gamma_{\mu}\partial_{\nu}\psi -\partial_{\nu}\bar{\psi}\gamma_{\mu}\psi) 
+(\mu\leftrightarrow \nu)\right)-\eta_{\mu\nu}\mathcal{L}_0
\eeq
that we couple to the external field $h^{ext}_{\mu\nu}$. In momentum space the corresponding vertex takes the 
form (see Fig. \ref{FF})
\beq
V^{(0) \mu\nu}=\frac{i}{4}\left( \gamma^\mu(p_1 +p_2)^\nu + \gamma^\nu(p_1 +p_2)^\mu -
2 \eta^{\mu\nu}(\slashed{p}_1+\slashed{p}_2 - 2 m)\right),
\eeq
\begin{figure}[t]
\centering
\subfigure[]{\includegraphics[scale=1]{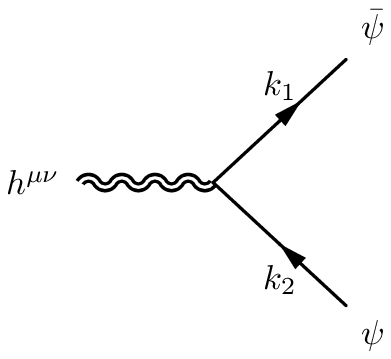}} \hspace{3cm}
\subfigure[]{\includegraphics[scale=1]{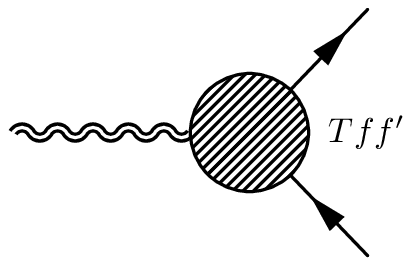}} \hspace{.5cm}
\caption{The leading order (a) graviton/fermion/fermion vertex. Fig. (b) symbolizes the radiative corrections.   }
\end{figure}
from which we have omitted an overall factor $(-\kappa/2)$ coming from the interaction Lagrangian, which will be reinserted at the level of the transition amplitude. Only diagonal transition amplitudes in flavour space are induced at leading order, while the inclusion of the electroweak corrections allows flavour-changing transitions, which will play a key role in our analysis.

We will denote with $\hat{T}^{\mu\nu}$ the transition amplitude between the intial and the final state induced 
by the insertions of the EMT vertex, with initial and final wave functions included, while $V^{\mu\nu}$ will be denoting the corresponding vertex in momentum space, i.e. 
\beq
\hat{T}^{\mu\nu}=\bar u(p_2)V^{\mu\nu}u(p_1),
\eeq
having labeled the momenta of the incoming and outgoing fermion with $p_1$ and $p_2$ respectively.  
We will similarly use the notation 
\beq
\hat{O}\equiv \bar{u}(p_2) O u(p_1)
\eeq
to denote the matrix element of any operator $O$ in momentum space.   
Using these notations,  the same matrix element can be expressed in position space in terms of the operator $T^{\mu\nu}(x)$, inserted on the Dirac eigenstates given by incoming ($\psi_i$) and outgoing ($\psi_f$) plane waves
\beq
\langle p_2|T^{\mu\nu}(x)|p_1\rangle=\bar{\psi_f}(p_2)V^{\mu\nu}\psi_i(p_1) e^{i q\cdot x},
\eeq
where $q=p_1-p_2$ is the 4-momentum transfer. We have introduced plane waves given by 
\beq
\psi_{i}(p_{1})={\mathcal{N}_{i} }u(p_{1}), \qquad \mathcal{N}_{i}=\sqrt{\frac{m_{1}}{E_{1} V_l}}, \qquad \bar{u}(p_1)u(p_1)=1,
\eeq
and similarly for $\psi_f$,  while $V_l$ denotes a finite volume normalization of the two scattering states. The $E_1$ ($E_2$) and $m_1$ ($m_2$) are the energy and corresponding mass of the incoming (outgoing) particle respectively.

The scattering matrix element is then written as 
\beq
i\mathcal{S}_{if}=-\frac{\kappa}{2}\int_{\mathcal V} d^4 x \langle p_2 |h_{\mu\nu}(x) T^{\mu\nu}(x)| p_1 \rangle 
\label{volume},
\eeq
where ${\mathcal V}$ denotes the relevant region of integration and with the gravitational field being external, which gives
\bea
 \langle p_2 |h_{\mu\nu}(x) T^{\mu\nu}(x)| p_1 \rangle&=& h_{\mu\nu}(x) \bar{\psi}(p_2)V^{\mu\nu}\psi(p_1) e^{i q\cdot x}.
  \eea
The scattering amplitude which can be immediately expressed in momentum space as
 \bea
i \mathcal{S}_{fi}&=& -\frac{\kappa}{2} h_{\mu\nu}(q) \bar{\psi}(p_2)V^{\mu\nu}\psi(p_1)\nn \\
&=&  -\frac{\kappa}{2}  h_{\mu\nu}(q)  \mathcal{N}_i\mathcal{N}_f \hat{T}^{\mu\nu}
\label{sfi}
\eea
in terms of the gravitational fluctuations in momentum space $h_{\mu\nu}(q)$, defined as in (\ref{h2}).\\
For a static 
external field the energies of the incoming/outgoing fermions are conserved ($E_1=E_2\equiv E$) both in the flavour-diagonal and non-diagonal cases, as one can immediately realize from (\ref{h1}). It is obvious that the interaction region depends both on the type of scattering particle and on the geometry of the source. 
We will consider the potential scattering of a Dirac fermion off an external static source, which acts as a 
perturbation on the otherwise flat spacetime background. The source is characterized by an EMT $T^{ext}_{\mu\nu}$ and the fluctuations are determined by solving the linearized equations of General Relativity. 

 \section{Leading order cross section} 

We start analyzing the flavour-conserving transitions at leading order. In the rest of the paper, for simplicity, we will omit the flavour index from each spinor, together with the explicit flavour dependence on the masses of the internal flavours, which appear at 1-loop order. The linearized equations take the form
\beq
\square\left(h_{\mu\nu}-\frac{1}{2}\eta_{\mu\nu} h\right)=-{\kappa}T^{ext}_{\mu\nu}
\eeq
and can be rewritten as
\beq
\square h_{\mu\nu}={\kappa} S_{\mu\nu}, \qquad S_{\mu\nu}=-\left( T^{ext}_{\mu\nu}-\frac{1}{2}\eta_{\mu\nu}T^{ext}\right) 
\label{oneq},
\eeq
where we have denoted with $T^{ext}$ the EMT trace of the external source.  
The external field is obtained by convoluting the static source with the retarded propagator 
\beq
G_R(x,y)=\frac{1}{4 \pi}\frac{\delta(x_0-|\vec{x}-\vec{y}|-y_0)}{|\vec{x}-\vec{y}|}
\eeq
normalized as
\beq
\square G_R(x,y)=\delta^4(x-y). 
\eeq
The solution of (\ref{oneq}) takes the form 
\beq
h^{ext}_{\mu\nu }(x)={\kappa}\int d^4 y G_R(x,y)S_{\mu\nu}(y),
\eeq
with the EMT of the external localized source 
\beq
T^{ext}_{\mu\nu}=\frac{P_{\mu}P_{\nu}}{P_0}\delta^3(\vec{x}), 
\eeq
which for a particle  of mass $M$ at rest at the origin with $P_{\mu}=(M,\vec{0}$) takes the form
\beq
\label{emt}
T^{ext}_{\mu\nu}=M\delta^0_\mu\delta^0_\nu\delta^3(\vec{x}).
\eeq
This gives 
\beq
S_{\mu\nu}=\frac{M}{2}\bar{S}_{\mu\nu} \qquad \bar{S}_{\mu\nu}\equiv \eta_{\mu\nu}-2 \delta^0_{\mu}\delta^0_{\nu}
\eeq
and the field generated by a local (point-like, $L$) mass distribution has a typical $1/r$  $(r\equiv |\vec{x}|)$ behaviour
\bea
h^L_{\mu\nu}(x)
&=& \frac{2 G M}{\kappa |\vec{x}|}\bar{S}_{\mu\nu}.
\label{hh}
\eea
The fluctuations are normalized in such a way that $h_{\mu\nu}$ has mass dimension 1, as an ordinary bosonic field, with 
$\kappa$ of mass dimension ${-1}$. 

The Fourier transform of $h_{\mu\nu} $ in momentum space is given by 
\bea
 h_{\mu\nu}(q_0,\vec{q})&=&\int d^4 x e^{i q\cdot x} h_{\mu\nu}(x)\nn\\
 \eea
 which for a static field reduces to the form 
 \beq
 h_{\mu\nu}(q_0,\vec{q})=2 \pi \delta(q_0) h_{\mu\nu}(\vec{q} ),
 \label{h1}
 \eeq
 with 
 \beq
 h_{\mu\nu}(\vec{q})\equiv h_0(\vec{q}) \bar{S}_{\mu\nu}
\label{h2}
 \eeq
 defined in terms of the scalar form factor  which is related to the geometrical structure of the external source.  
In the case of a point-like source, we obtain 
 \bea
h_{\mu\nu}(q_0,\vec{q})&=& 2 \pi \delta(q_0)\times \frac{2 G M}{\kappa} \bar{S}_{\mu\nu}\int d^3 \vec{x} \frac{e^{i\vec{q}\cdot \vec{x}}}{|\vec{x}|}\nn\\
&=& 2 \pi \delta(q_0)\times \left(\frac{\kappa M}{2\vec{q}^2}\right)\bar{S}_{\mu\nu},
\label{trans}
\eea
which gives
\beq
 h_0(\vec{q})\equiv \left(\frac{\kappa M}{2 \vec{q}^2}\right), \qquad  h_{\mu\nu}(\vec{q})\equiv \left(\frac{\kappa M}{2 \vec{q}^2}\right) \bar{S}_{\mu\nu}.
 \label{h0}
 \eeq
At the same time, the leading order interaction is redefined as 
\beq
V^{(0) \mu\nu}\equiv{i}O_V^{(0) \mu\nu},
\eeq
with 
\beq
O_V^{(0) \mu\nu}=\frac{1}{4}\left( p^\mu\gamma^\nu + p^\nu\gamma^\mu - 2 \eta^{\mu\nu}(\slashed{p} - 2 m)\right),
\eeq
which will turn useful in the computation of the radiative corrections.
Using the matrix element 
\beq
\bar{\psi}_f(x) T^{(0)\mu\nu}\psi_i(x) =\left(-\frac{\kappa}{2}\right)\times \mathcal{N}_i\mathcal{N}_f\times i \bar{u}(p_2)O_V^{(0) \mu\nu}u(p_1) e^{i q\cdot x} 
\eeq
the leading order scattering amplitude can be rewritten as
\bea
i\mathcal{S}^{(0)}_{i f}&=&\int d^4 x h_{\mu\nu}(x)\bar{\psi}_f(x) T^{(0)\mu\nu}\psi_i(x) \nn\\
&=& \left(-\frac{\kappa}{2}\right)\times \mathcal{N}_i\mathcal{N}_f\times\left(i h_{\mu\nu}(\vec{q})\bar{u}(p_2)O_V^{(0) \mu\nu}u(p_1)\right) \times 2\pi\delta(q_0).
\eea
The averaged squared amplitude then takes the form 
\bea
\langle|i\mathcal{S}^{(0)}_{i f}|^2\rangle&=&\left(-\frac{\kappa}{2}\right)^2 \left(\mathcal{N}_i
\mathcal{N}_f\right)^2\times  \left(2\pi\delta(q_0) \mathcal{T}\right)\times \left(\frac{\kappa M}{2\vec{q}^2}\right)^2 \times\frac{1}{2}\mathcal{Y}_0,
\label{averaged}
\eea
where we have introduced a factor (1/2) for the fermion spin average. As usual, we have extracted, from the square of the delta function, the transition time $\mathcal{T}$, using  $(2\pi\delta(q_0))^2=2\pi\delta(q_0)\mathcal{T}$. We have defined
\bea
\mathcal{Y}_0&=&\frac{1}{4 m^2}Tr\left[ (\slashed p_2 + m) O_V^{(0) \mu\nu} (\slashed p_1 + m)O_V^{(0)\alpha\beta}\right]\bar{S}_{\mu\nu} \bar{S}_{\alpha\beta} \nn\\
&=&
E^2\textrm{Tr}\left[\frac{\slashed{p_2}+m}{2 m}\left(2 \gamma^0-\frac{m}{E}\right) \frac{\slashed{p_1}+m}{2 m}\left(2 \gamma^0-\frac{m}{E}\right)\right]\nn \\
&=&
8\frac{\vec{p_1}^4}{ m^2 }F^{(0)}(x,\theta)
\eea
in the flavour-diagonal case, with $F^{(0)}(x,\theta)$ given by 
\beq
F^{(0)}(x,\theta)=\cos^2\frac{\theta}{2} + \frac{x}{4} 
+ \frac{x^2}{4} + \frac{3}{4}{x}\cos^2\frac{\theta}{2},
\label{given}
\eeq
with $x=m^2/\vec{p_1}^2$ and $\vec{p_1}$ the 3-momentum of the incoming fermion. We have used the relations $ \vec{p}_f\equiv|\vec{p}_1|=|\vec{p}_2|$ in the elastic limit, while $\vec{q}^{\,2}=4 \vec{p_1}^{\,2} \sin^2\theta/2$ and $q^2=-\vec{q}^{\,2}$.

To compute the differential cross section we define the transition probability from the initial state ($i$) into a set of final states ($f$) of differential $d n_f$ given by 
\beq
d W= \frac{\vline\, i\mathcal{S}_{if}\,\vline^2}{j_i} dn_f
\eeq
normalized with respect to the incident flux $j_i=\vec{p_1}/(E_i V)$
with the final state density, with a volume normalization $V$ given by 
\beq
d n_f=\frac{V}{(2 \pi)^3} d^3 \vec{p}_f=\frac{V}{(2 \pi)^3}|\vec{p_2}|E_2 dE_2 d\Omega,
\eeq
which allows to define the cross section as the differential transition rate per unit time $(d\sigma\equiv{d W}/{\mathcal{T} })$.
Combining all the contributions together and integrating over the energy of the final state we obtain 
\bea
\frac{d \sigma}{d \Omega}& \equiv& \frac{d \sigma}{d \Omega}|_L\nn\\
&=&\left(\frac{G M }{\sin^2\frac{\theta}{2}}\right)^2 F^{(0)}(x,\theta).
\label{leading}
\eea
The result in Eq.~(\ref{leading}) is in agreement with the corresponding
one evaluated in Ref. \cite{Lawrence:1970fx}.
 In the following sections we will refer to this cross section 
as to the {\em localized} one, denoted in (\ref{leading}) also as $d\sigma/d\Omega|_L$, being generated by a gravitational point-like source located at the origin. 
This will turn useful in order to make a distinction between the cross sections of localized and of extended sources.
The computation of the angle of deflection for the fermion involves a simple semiclassical analysis, in which one introduces the 
impact parameter representation of the same cross section. 
Assuming that the incoming particle is moving along the $z$ direction, with the source localized at the origin, and denoting with $\theta$ the azymuthal angle,  
we have the relation
\beq
\frac{b}{\sin\theta}\vline\frac{d b}{d\theta}\vline=\frac{d \sigma}{d\Omega}
\label{semic}
\eeq
between the impact parameter $b$ and the scattering angle $\theta$, measured from the $z-$direction. This semiclassical approach allows to relate the quantum interaction between the particle and the source. We recall, if not obvious, that in the classical description the relation between $b$ and the deflection angle $\theta_d$ is obtained from energy and angular momentum conservation. In this respect Eq.~(\ref{semic}) takes a similar role. The relation between $b$ and $\theta$ as a function of the incoming energy ($b=b(E,\theta)$), at least for small deflection angles, which correspond to large impact parameters, can be found analytically, but it should be integrated numerically otherwise.   
In the case of the point-like cross section $(L)$, for instance, one obtains the differential relation 
\beq
\frac{d b^2}{d\theta}=- 2  \left(\frac{G M }{\sin^2\frac{\theta}{2}}\right)^2 F^{(0)}(x,\theta) \sin\theta
\eeq
which gives 
\beq
b^2(\theta)=(G M)^2 \left( \frac{(2 + x)^2}{\sin^2(\frac{\theta}{2})} + 2 (4 + 3 x )\log\left( \sin(\frac{\theta}{2})\right)\right).
\eeq
In the small $\theta$ limit we get the relations 
\beq
b \sim G M\left(\frac{4}{\theta} +  \frac{2 x}{\theta} + ( 1 + \frac{x}{4})\theta \log\theta\right) + {\cal O}(x^2 \theta\log \theta) + 
{\cal O}(\theta),
\label{blocal}
\eeq
which allows us to identify the deflection angle as 
\beq
\theta\equiv\theta_d \sim 4 \frac{G M}{b}.
\label{impact}
\eeq
Notice that in general $x \ll 1$ for a scattering at high energy, being the fermion mass small compared to the incoming energy or three-momentum.  Onbviously, the bending angle of neutrinos, charged leptons and dark matter particles will be affected by the geometry of the source and by the strength of the fermion
interaction with the source. This will vary if, for instance, the particles are allowed to enter the core of the gravitational source, as in the neutrino and the dark matter cases.  For this reason we will proceed by determining the structure of these corrections, providing more general cross sections beyond the point-like approximation.  

\subsection{Modified leading order cross sections} 
 Once we allow a finite size for the mass distribution, expression (\ref{leading})  gets modified. This is the case, for instance, for beams of neutrinos or dark matter particles which may have, in principle, impact parameters smaller than the source radius. In this case, it is intuitively clear that in the limit of a weak gravitational background, Gauss' theorem in gravity implies that only the region of the source of radius smaller than the particle impact parameter is relevant in the scattering process. This can be shown rigorously directly from the spherically symmetric Schwarzschild metric. However, the proof - in the weak field limit - of factorization of the 3-dimensional euclidean metric from the Schwarzschild solution, requires some technical steps which are left to an appendix (appendix \ref{Schw}). This allows to avoid the use of polar projectors in the scattering amplitude - for spherically symmetric backgrounds - which would render the treatment more involved from the computational side. 
Our discussion, here, is limited to the case of a uniform mass distribution, but it can be generalized to any 
distribution of spherical symmetry, in the weak field limit.  Using this result, it follows that the metric fluctuations of a distributed source can be described (in $c=1$ units) by the field 
\beq
h_{\mu\nu}= - \frac{2 \Phi_d}{\kappa}\bar{S}_{\mu\nu}
\eeq
with the potential of the distributed source $\Phi_d$, given by 
 \beq
 \Phi_d(r)=-\frac{G M}{r'}\theta(r - R) -\frac{G M}{2 R}(3 - \frac{r^2}{R^2})\theta(R-r).
 \label{pot}
 \eeq
Notice that in the expression above we have separated the contribution of the external core region from the internal one, with $R$ being the radius of the source. The modifications to the expression of the cross section in the various cases, due to the different nature of the scattering particles, can be taken into account by the insertion of the appropriate form of the fluctuation tensor in momentum space, obtained from this generalized geometrical setting.
\begin{itemize} 
\item{\bf Non-weakly interacting fermions}
\end{itemize}
For this reason, as in the case of the localized gravitational source, for non-weakly interacting fermions we restrict 
the integration region of the scattering amplitude in the form 

\beq
i\mathcal{S}_{if}=-\frac{\kappa}{2}\int d x_0\int_{>R} d^3\vec{x} \langle p_2 |h_{\mu\nu}(x) T^{\mu\nu}(x)| p_1 \rangle, 
\eeq
where we have restricted the integration region to $|\vec{x}|>R$, i.e. outside the radius of the gravitational source. Therefore, according to (\ref{pot}), it is convenient to denote the contributions to the gravitational field coming from the external region as $h_{E \mu\nu}$. At leading order, for interactions which involve only the region of the external core  we have

 \bea
i \mathcal{S}_{fi}&=&  -\frac{\kappa}{2}\,  h_{E\mu\nu}(q_0,\vec{q})  \mathcal{N}_i\mathcal{N}_f \hat{T}^{\mu\nu}
\label{sfi}
\eea
with  
\bea
h_{E\mu\nu}(q_0,\vec{q})&=& 2\pi\delta(q_0) h_{E}(\vec{q})\bar{S}_{\mu\nu}, \nn\\
h_{E}(\vec{q})&=& \int_{>R} d^3 \vec{x} \left( - \frac{2 \Phi_d}{\kappa}\right) e^{i\vec{q}\cdot \vec{x}} \nn\\
&=&\frac{\kappa M}{2 \vec{q}^2}\cos( |\vec{q}| R)
\label{out}
\eea
obtained as Fourier transform of the potential in the restricted region. It is then clear that, if we allow only interactions out of the core region, as we should 
for non-weakly interacting fermions, the modification of the cross section and of the expression of the bending angle, at leading order, are given by  
\beq
\frac{d \sigma}{d \Omega}\vert_{E}=\cos^2( |\vec{q}| R) \frac{d \sigma}{d \Omega}|_L,
\label{sigmaM}
\eeq
having used for $d\sigma/d\Omega|_L$  the result given in (\ref{leading}).   
\begin{itemize} 
\item{\bf Neutrinos and dark matter fermions} 
\end{itemize} 
Moving to the case of neutrino/ dark matter scatterings, as in the previous cases, the interaction region covers all the 3-dimensional space. Beside $h_E^{\mu\nu}(x)$, it is convenient to introduce also the expression of the field generated inside the core of the source $h_I^{\mu\nu}(x)$ and its corresponding form factor $h_I(q)$. 
In this case the toal form factor $h(\vec{q})$ is given by 
\beq
h(\vec{q})= h_I(\vec{q}) + h_E(\vec{q})
\label{all}
\eeq
with 
\bea
 h_I(\vec{q})&=&\int_{< R}d^3\vec{x}\left(- \frac{2\Phi_D}{\kappa} \right) e^{i\vec{q}\cdot \vec{x}} \nn\\
 &=& -\frac{8 \pi  {G M} \left(|\vec{q}| R \left(\vec{q}^{\,2}
   R^2+3\right) \cos (|\vec{q}| R)-3 
\sin (|\vec{q}| R)\right)}{\kappa  |\vec{q}|^5 R^3}
\eea
while $h_E$ is given by (\ref{out}). Using (\ref{all}), the gravitational field in momentum space is given by 
\beq
h_{\mu\nu}(q_0,\vec{q})= -\frac{24\pi G M}{\kappa |\vec{q}|^5 R^3}\bigg(   |\vec{q}| R \cos( |\vec{q}|R) - \sin( |\vec{q}| R)\bigg)\bar{S}_{\mu\nu}\times 2\pi \delta(q_0)
\eeq
(in units $c=1$).
The corresponding cross section at leading order is proportional to the result for the localized one 
\beq
\frac{d \sigma}{d \Omega}\vert_{g}=A(\vec{q})^2 \frac{d \sigma}{d \Omega}|_L
\label{distr}
\eeq
with the modulating factor given by 
\beq
A(\vec{q})= -\frac{3}{|\vec{q}|^3 R^3}\left( |\vec{q}| R \cos( |\vec{q}|R)-\sin( |\vec{q}|R)\right).
\label{distr1}
\eeq
 
 \begin{figure}[t]
\centering
\subfigure[]{\includegraphics[scale=1]{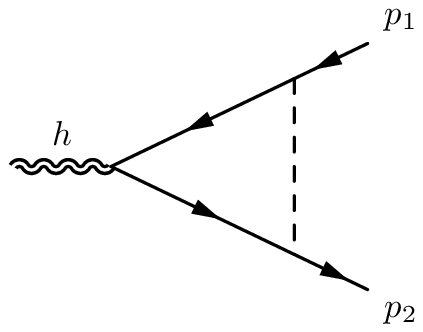}} \hspace{3cm}
\subfigure[]{\includegraphics[scale=1]{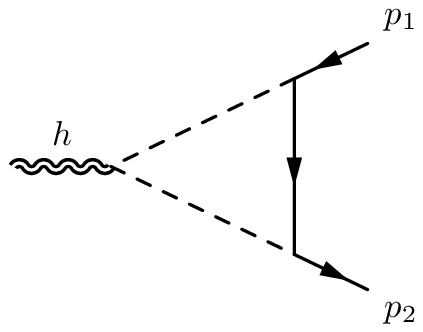}} \hspace{.5cm}
\subfigure[]{\includegraphics[scale=1]{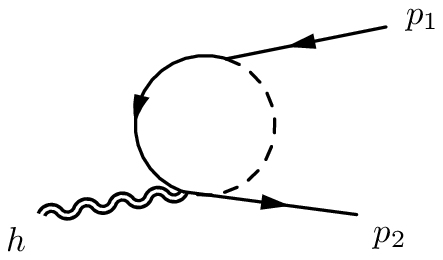}} \hspace{.5cm}
\caption{Typical topologies of vertex corrections included in the computation of the radiatve corrections in the flavour-diagonal and non-diagonal cases. }
\end{figure}
 
\section{The gravitational fermion form factors at 1-loop: { definitions and conventions}}
The one-loop corrections to the graviational fermion  vertex in the framework of the SM, have been computed in Refs. \cite{Coriano:2012cr} and \cite{Degrassi:2008mw,Coriano:2013msa} 
for the flavor diagonal and off-diagonal transitions, for the case
of external on-shell fermions. Exact results at one-loop order in all masses and momenta are provided for the corresponding form factors in Refs. \cite{Coriano:2012cr} and \cite{Coriano:2013msa}. \\
The expression up to 1-loop order of the contributions to the fermion-fermion-graviton vertex $(ffg)$ can be organized in terms of a flavour-conserving (or diagonal) sector contribution and of a flavour-changing part. The diagonal sector receives contributions at tree level  (i.e. from $\hat{T}^{(0) }_{\mu\nu }$), and at 1-loop from the virtual exchange of a photon 
($\hat T_{\gamma}^{\mu\nu}$), a gluon ($\hat T_{g}^{\mu\nu}$), a $Z$ 
($\hat T_{Z}^{\mu\nu}$) and a $W$, the latter denoted as $\hat T^{D\, \mu\nu}_{W}$. 
In the flavour-changing (non diagonal) sector the corrections start at 1-loop and are generated only from the exchange of a virtual $W$, denoted as 
$\hat T^{ND\, \mu\nu}_{W, {\bf ff'}}$.
As usual, the radiative corrections can be splitted into vertex corrections and the related
counter-terms. We use the expression {\em counter-terms} to refer to
the sum of those diagrams coming from the self-energy contributions to the external 
fermion legs, plus the corresponding diagrams coming from their wave-function renormalization. In the case of the flavor-changing transitions, the latter contributions are absent at one-loop level in the EMT counterterms, due to the fact that the 
fermion-graviton vertex is flavor-conserving at tree-level \cite{Degrassi:2008mw,Coriano:2013msa}.
Because of the energy momentum conservation, the radiative corrections induced by SM fields (no graviton loop corrections) to the EMT are finite at any order in perturbation theory, which implies the nonrenormalization 
of the Newton constant.

It is convenient to separate the diagonal ($D$) from the non-diagonal ($ND$) parts using the notation
\beq
T^{\,\mu\nu}_{\bf f f'}= T^{D,\mu\nu}_{\bf ff'} + T^{ND\,\mu\nu}_{\bf ff'}
\eeq
with $T^{D\,\mu\nu}_{\bf ff'}$ expressed as  
\beq
\label{diag}
T^{D\,\mu\nu}_{\bf f f'}=\delta_{\bf f f'} T^{D\,\mu\nu},
\eeq
where the flavour-conserving transitions are
\beq
\hat T^{D\,\mu\nu}= \hat{T}^{(0) }_{\mu\nu } + \hat T^{\mu\nu}_{g} + \hat T^{\mu\nu}_{\gamma} + \hat T^{\mu\nu}_{h} + \hat T^{\mu\nu}_{Z}
\eeq
while the flavour-changing matrix elements take the form
\bea
\label{hatT}
\hat T^{ND\, \mu\nu}_{\bf f f'} = \hat T^{ ND\,\mu\nu}_{W,\bf  ff' } 
\eea
since these can be generated only by the W exchange.
The contributions given above are expanded in terms of a set of operators $O_l$ and intermediate form factors 
$F_l$, multiplied by the CKM matrix elements, related to internal the flavour transitions in the internal vertices of the diagrams. These are accounted for by the $V_{\bf fn}$ matrix elements, with the indices $\bf f$ and $\bf n$ 
corresponding to the flavour of the external and internal-loop fermions
respectively. Generically, the flavour structure can be completely absorbed into some final form factors $f^{w}_l(q)$    
\bea
\label{Ftof}
f^{w \,\bf{f f'}}_l(q^2) =  \sum_{\bf n} V_{\bf f n}^* V_{\bf n f'} \, F_l(q^2, x_f), 
\label{fw}
\eea
as we are going to specify below.\\
In the followings we will drop the flavour dependence on $f^{w}_l$, just to simplify the expressions, and the distinction between the (flavour) diagonal and non diagonal cases will be made evident by use of the superscript $ND$  (Non Diagonal) in the flavour-changing cases. In the diagonal case the contributions coming from the electromagnetic and weak transitions will be summarized by the form factors $f_k$, with $k=1\ldots 4$ and $(f^h_l, f^Z_l, f^W_l,)$ with $l=1,\ldots 6$ respectively. In the non diagonal case, the $f^{ND}_j$, with $j=1,\ldots 12$, will completely describe all the corresponding transitions. The flavour structure will be omitted also from the EMT in both cases. We will also denote with $m_{f},m_{f'}$ the masses of the initial and final state fermions. All the expressions of the form factors discussed in this work can be found in \cite{Degrassi:2008mw, Coriano:2012cr, Coriano:2013msa}.
\subsection{ Diagonal contributions}
In the diagonal sector, vector-like ($V$) matrix elements are expanded onto a basis of four tensor operators $O^{\mu\nu}_{V  k}$ as
\bea
\label{vectorbasis}
O^{\mu\nu}_{V  1} &=& \gamma^\mu \, p^\nu + \gamma^\nu \, p^\mu \,, \nn \\
O^{\mu\nu}_{V 2} &=& m \, \eta^{\mu\nu} \,, \nn \\
O^{\mu\nu}_{V  3} &=& m \, p^\mu \, p^\nu \,, \nn \\
O^{\mu\nu}_{V 4} &=& m \, q^\mu \, q^\nu, \,
\eea 
where $m$ is the mass of the corresponding fermion,
generated by the photon, gluon and Higgs exchanges.

A second set of operators chiral operators $(C)$ appears in the exchange of $W$'s and $Z$ gauge bosons in the virtual corrections
\bea
\label{chiralbasis}
O^{\mu\nu}_{C  1} &=& \left( \gamma^\mu \, p^\nu + \gamma^\nu \, p^\mu \right) P_L \,, \nn \\
O^{\mu\nu}_{C  2} &=& \left( \gamma^\mu \, p^\nu + \gamma^\nu \, p^\mu \right) P_R \,, \nn \\
O^{\mu\nu}_{C  3} &=& m \, \eta^{\mu\nu} \,, \nn \\
O^{\mu\nu}_{C  4} &=& m \, p^\mu \, p^\nu  \,, \nn \\
O^{\mu\nu}_{C  5} &=& m \, q^\mu \, q^\nu  \,, \nn \\
O^{\mu\nu}_{C  6} &=& m \, \left( p^\mu \, q^\nu + q^\mu \, p^\nu \right) \gamma^5, \,
\eea
where $P_{L,R}=(1\mp \gamma_5)/2$, whose contributions are summarized by the amplitudes 
\beq
\label{akin}
\hat{O}^{\mu\nu}_{C l}\equiv \bar{u}(p_2)O^{\mu\nu}_{C l}u(p_1).
\eeq
 We have imposed the  symmetry constraints on the external fermion states (of equal mass and flavour) and the conservation of the EMT. Explicitly, the various contributions are given by 

\bea
\hat T^{\mu\nu}_{g} &=& i \frac{\alpha_s }{4 \pi} C_F  \sum_{k=1}^4 f_k(q^2) \, \bar u(p_2) \, O^{\mu\nu}_{V  k} \, u(p_1) \,, \\
\hat T^{\mu\nu}_{\gamma} &=& i \frac{\alpha }{4 \pi} Q^2  \sum_{k=1}^4 f_k(q^2) \, \bar u(p_2) \, O^{\mu\nu}_{V k} \, u(p_1) \,, \\
\hat T^{\mu\nu}_h &=& i  \frac{G_F}{16 \pi^2 \sqrt{2}}  \, m^2  \sum_{k=1}^4 f^h_k(q^2) \, \bar u(p_2) \, O^{\mu\nu}_{V  k} \, u(p_1) \,,\eea
with $C_F=(N_c^2 -1)/(2 N_c)$ and $N_c=3$ for QCD. 
The remaining diagonal contributions, generated by the virtual exchange of a $Z$ or of the $W$'s gauge bosons, are expanded on a tensor basis of six elements as
\bea
\hat T^{\mu\nu}_Z &=& i \, \frac{G_F}{16 \pi^2 \sqrt{2}}   \sum_{k=1}^6  f^{Z}_k(q^2) \, \bar u(p_2) \, O^{\mu\nu}_{C  k} \, u(p_1) \,, \\ 
\hat T^{\mu\nu}_W &=& i \, \frac{G_F}{16 \pi^2 \sqrt{2}}  \sum_{k=1}^6  f^{W}_k(q^2) \, \bar u(p_2) \, O^{\mu\nu}_{C  k} \, u(p_1). 
\eea
The exact expressions
for the form factors $f_k(q^2)$, $f_k^{h,W,Z}(q^2)$ appearing above, can be found in Ref.\cite{Coriano:2012cr}.
For convenience we introduce the linear combinations 
\beq
\mathcal{R}^{(a)}_k= i\left(\frac{\alpha_s}{2\pi} C_F +\frac{\alpha}{4 \pi} Q^2\right) f_k(q) + 
i\frac{G_F}{16 \pi^2 \sqrt{2}} m^2 f_k^h(q) \qquad k=1,2,\ldots 4
\label{Ra}
\eeq
where the form factors $f_k(q)$ are those generated by the exchange of a photon or a gluon, while the 
$f^h_k(q)$ are those due to a virtual Higgs. The chiral operators, instead, multiply the linear combinations 
\beq
\mathcal{R}^{(b)}_l=  i\frac{G_F}{16 \pi^2 \sqrt{2}} \Big(f^Z_l(q) +f_l^W(q)\Big) \qquad l=1,2,\ldots 6
\label{Rb}
\eeq
where $f^Z_l(q)$ and $f_l^W(q)$ denote the $W$'s and $Z$ form factors, contributing to the flavour-diagonal case. In Eq.~(\ref{fw}) we have set $x_f\equiv m_f^2/m_W^2$, where $m_f$ and $m_W$ stand for
the masses of the fermion $f$ and of the $W$ gauge bosons respectively. In the following we will be using the short-hand 
notation $\mathcal{R}_i$ to denote the 10 form factors above  ($\mathcal{R}_i\equiv(\mathcal{R}^{(a)}_k,  \mathcal{R}^{(b)}_l))$.

Notice that in the case of an elastic scattering of a neutrino, only $O_{V 1}, O_{C 1}$ and $O_{C2}$ contribute to the cross section.

\subsection{ Flavour-changing contributions} 
 As we have mentioned above, the flavour-changing contributions appear at 1-loop order induced by the W-boson exchange. They include the vertex corrections plus the corresponding flavor-changing counter terms, that consists of
the flavor-changing self-energy insertions on the external fermion legs
\cite{Degrassi:2008mw,Coriano:2013msa}.
The total contribution in this sector can be expressed in terms of a set of operators $O_k$ as 
\bea
\label{hatT}
\hat T^{ND \,\mu\nu}_{W} = -i\frac{G_F}{16 \pi^2 \sqrt{2}} \sum_{k=1}^{12} f^{ND}_k(q^2) \, \bar u^{}(p_2) O^{\mu\nu}_k u^{}(p_1)
\label{TFC}
\eea
with  the tensor operator basis given by
\bea
\begin{array}{ll}
O_1^{\mu\nu} = \left( \gamma^{\mu} p^{\nu}+\gamma^{\nu} p^{\mu}\right) P_L \qquad \qquad
& O_7^{\mu\nu}=  \eta^{\mu\nu} \, M_- \\
O_2^{\mu\nu}= \left( \gamma^{\mu} q^{\nu}+\gamma^{\nu} q^{\mu}\right) P_L 
& O_8^{\mu\nu}=  p^{\mu}p^{\nu} \, M_-  \\
 O_3^{\mu\nu}= \eta^{\mu\nu} \, M_+  
&O_9^{\mu\nu}=  q^{\mu}q^{\nu} \, M_-  \\
O_4^{\mu\nu}= p^{\mu}p^{\nu} \, M_+
&O_{10}^{\mu\nu}= \left(p^{\mu}q^{\nu}+q^{\mu}p^{\nu}\right) M_-\\
O_5^{\mu\nu}=  q^{\mu}q^{\nu} \, M_+
& O_{11}^{\mu\nu}= \frac{m_i m_j}{m_W^2}  \left( \gamma^{\mu} p^{\nu}+\gamma^{\nu} p^{\mu}\right) P_R \\
O_6^{\mu\nu}=  \left(p^{\mu}q^{\nu}+q^{\mu}p^{\nu}\right)\, M_+
&O_{12}^{\mu\nu}= \frac{m_i m_j}{m_W^2}  \left( \gamma^{\mu} q^{\nu}+\gamma^{\nu} q^{\mu}\right) P_R, 
\end{array}
\label{basis}
\eea
where $M_{\pm}\equiv m_j P_R\pm m_i P_L$.
This is the most general rank-2 tensor basis that can be built out of two momenta, $p$ and $q$, a metric tensor and Dirac matrices $\gamma^\mu$ and $\gamma^5$ \cite{Degrassi:2008mw}. \begin{figure}
\centering
\subfigure[]{\includegraphics[scale=1]{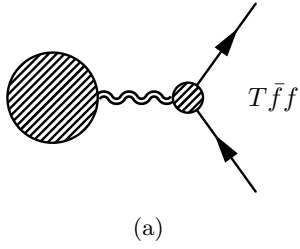}} \hspace{3cm}
\caption{Fermion scattering off an external source in the static limit.}
\end{figure}
Notice that in the equal mass case $(m_i=m_j)$ $M_+$ is proportional to the identity matrix and $M_-$ is a pseudoscalar 
contribution. In the massless case, relevant for neutrino flavour transitions, only $O_1$ and $O_2$ survive. The exact analytical expressions for the 
form factors $f^{ND}_k(q^2)$, with on-shell fermions, 
have been computed in \cite{Coriano:2013msa} and
at the leading order in $q^2$ in \cite{Degrassi:2008mw}.

\section{Next to leading order cross sections}
\subsection{Flavour-diagonal cross section at 1 loop}
Using these notations, the scattering matrix element, in the diagonal case, can be expressed in the form 
\bea
\label{diagS}
i\mathcal{S}_{i f}&=& -\frac{\kappa}{2}\mathcal{N}_i\mathcal{N}_f h_{\mu\nu}\hat{T}^{D}_{\mu\nu} \nn \\
&=&-\frac{\kappa}{2}\mathcal{N}_i\mathcal{N}_f h_{\mu\nu}\left( \hat{T}^{(0)}_{\mu\nu} + \sum_{k=1}^4 \mathcal{ R}^{(a)}_{k}\hat{O}^{\mu\nu}_{V k}  +
\sum_{l=1}^6 \mathcal{ R}^{(b)}_{l}\hat{O}^{\mu\nu}_{C l}\right)
\eea
with $T^D_{\mu\nu}$ given in (\ref{diag}) and the terms
$\mathcal{ R}^{(a,b)}_{k}$ defined in Eqs.(\ref{Ra}),(\ref{Rb}).

By a simple redefinition of the operators, 
\beq
O_D^i\equiv O_{V i},  i=0,\ldots,4 \qquad O_D^i\equiv O_{C i-4}, i=5,\ldots,10
\eeq
\beq
R_D^i\equiv R_{ i}^{(a)},  i=0,\ldots,4 \qquad R_D^i\equiv R_{i-4}^{(b)}, i=5,\ldots,10
\eeq
and defining 
\beq
\mathcal{Y}= \sum_{i,j=0}^{10}\langle O_D^i O_D^{j \dagger} \rangle \mathcal{R}^i_D  \mathcal{R}^{\dagger j}_D
\eeq
with 
\beq
\langle {O}_D^i {O}_D^{j\dagger} \rangle \equiv \frac{1}{4 m^2}Tr\left[ (\slashed p_2 + m) O_D^{i \mu\nu} (\slashed p_1 + m)O_D^{j\alpha\beta \dagger}\right]\bar{S}_{\mu\nu} \bar{S}_{\alpha\beta} 
\eeq
the cross section takes the form 
\beq
\frac{d \sigma}{d \Omega}=\left(\frac{\alpha e_Q^2}{4 \pi} \right)
\frac{\kappa^2 m^2}{32 \pi^2} h_0(q)^2\mathcal{Y}
\eeq
with 
$h_0(\vec{q})$ defined in (\ref{h0}). By computing the expression for 
$\mathcal{Y}$ defined above, we obtain for the differential cross section
 \bea
\frac{d \sigma}{d \Omega}&=&\frac{(G M)^2}{\sin^4\frac{\theta}{2}}\left(F^{(0)}(E,m,\theta) + \frac{\alpha e_Q^2}{4 \pi}\, 2 \mathcal{R}e \,F_D^{(1)}(E,m,\theta) +  \frac{\alpha_S}{4 \pi}C_F \,2\mathcal{R}e\, F_D^{(1)}(E,m,\theta)\right. \nn\\
&&\left.\qquad \qquad \qquad + 
\frac{G_F}{16\pi^2 \sqrt{2}} \,2 \mathcal{R}e\, F_D^{(3)}(E,m,\theta)\right),  
 \label{crs}
 \eea
 where we have dropped the $O(\alpha_i^2)$ terms in all the couplings. We have chosen to present (\ref{crs}) given in terms of the energy of the incoming/outgoing particle $E$ ($E=E_1=E_2$) and of its mass $m$, beside the scattering angle $\theta$.  $F^{(0)}$, the leading order term, is given by Eq.~(\ref{given}), which takes the equivalent form as a function of $E, m$ and $\theta$
 
\beq
F^{(0)}(E,m,\theta) =\frac{4 \text{E}^4-3 \text{E}^2 m^2+\left(4 \text{E}^4-5 \text{E}^2 m^2+m^4\right) \cos
   (\theta )+m^4}{8 
   \left(\text{E}^2-m^2\right)^2}.
  \label{F0}
   \eeq
For the rest, we have denoted with $F_D^{(1)}$ and $F_D^{(3)}$ the electromagnetic and weak corrections. 
The strong corrections, in the quark case, differ from the electromagnetic ones just by the colour factor, proportional to the 
 Casimir of $SU(3)$,  $C_F$.
\begin{itemize} 
\item{\bf Electromagnetic corrections}
\end{itemize}
Coming to the the electromagnetic corrrections, these can be organized in the form
 \beq
 F_D^{(1)}(E,m, \theta)=\sum_{k=1}^4X_k(E,m,\theta) f_k 
 \eeq
 in terms of the four form factors $f_k(q^2)$ and coefficient functions $X_k$.
 The $f_i$'s have been given in \cite{Coriano:2012cr} and, being a function of
$q^2$, can be expressed as functions of the variables
 $E, m$ and $\theta$.  In the following we will omit the $q^2$ dependence in the form factors $f_k$. For the expressions of the $X_i$ term we obtain
 \beq
X_1(E,m,\theta)= \frac{\left(m^2-4 \text{E}^2\right) \sin
   ^2\left(\frac{\theta
   }{2}\right)}{(\text{E}^2-m^2)}+\frac{\left(m^2-2
   \text{E}^2\right)^2}{\left(\text{E}^2-m^2\right)^2}
 \eeq
 
 \bea
 X_2(E,m,\theta)&=&\frac{m^2 \sin ^2\left(\frac{\theta }{2}\right)}{2
   (\text{E}^2- m^2)}+\frac{m^4-2 \text{E}^2
   m^2}{2 \left(\text{E}^2-m^2\right)^2}\nn\\
X_3(E,m,\theta)&=&\frac{2\left( m^4-2 \text{E}^2 m^2\right) \sin
   ^2\left(\frac{\theta
   }{2}\right)}{(\text{E}^2-m^2)}+\frac{\left(m^3-2
   \text{E}^2
   m\right)^2}{\left(\text{E}^2-m^2\right)^2}+{m^2}
   \sin ^4\left(\frac{\theta }{2}\right)\nn\\
X_4(E,m,\theta)&=&\frac{\left(2 \text{E}^2 m^2-m^4\right) \sin
   ^2\left(\frac{\theta
   }{2}\right)}{(\text{E}^2-m^2)}-{m^2}\sin
   ^4\left(\frac{\theta }{2}\right)
\eea
\begin{itemize} 
\item{\bf Weak corrections}
\end{itemize}
 Finally, the general expression of the function $F_D^{(3)}$ which contains the electroweak corrections to the unpolarized cross section is given by
\bea
F_D^{(3)}(E,m,\theta)&=&{W_0( E,m,\theta) }\frac{2 E^2 - m^2}{2(E^2- m^2)^2}+W_1(E,m,\theta) \frac{\sin^2\left(\frac{\theta}{2}\right)}{2(E^2-m^2)} +W_2(E,m,\theta)\sin^4\left(\frac{\theta}{2}\right) \nn\\
\eea
where
\bea
W_0(E,m,\theta) &=&
{f_3^h} \left(4 {E}^2 m^4-2
   m^6\right)+{f_1^h} \left(4
   {E}^2 m^2-2
   m^4\right)+{f_5^W} \left(2
   {E}^2-m^2\right)+{f_6^W}
   \left(2
  {E}^2-m^2\right)\nn \\
  && +{f_8^W}
   \left(4 {E}^2 m^2-2 m^4\right)+4
   {E}^2 {f_8^Z} m^2+2
   {E}^2 {f_5^Z}+2{E}^2
   {f_6^Z}-{f_2^h}
   m^4-{f_7^W} m^2-2 {f_8^Z}
   m^4 \nn\\
&&   -{f_5^Z} m^2-{f_6^Z}
   m^2-{f_7^Z} m^2
   \eea

\bea
W_1(E,m,\theta) &=& {f_3^h} \left(4 m^6-8 {E}^2
   m^4\right)+{f_4^h} \left(4
   {E}^2 m^4-2
   m^6\right)+{f_1^h} \left(2 m^4-8
   {E}^2 m^2\right)+{f_5^W}
   \left(m^2-4
   {E}^2\right) \nn\\
 &&  +{f_6^W}
   \left(m^2-4
   {E}^2\right)+{f_8^W} \left(4
   m^4-8 {E}^2
   m^2\right)+{f_9^W} \left(4
   {E}^2 m^2-2 m^4\right)-8 {E}^2 {f_8^Z} m^2 \nn\\
 &&  +4
   {E}^2 {f_9^Z} m^2-4
   {E}^2 {f_5^Z}-4 {E}^2
   {f_6^Z}+{f_2^h}
   m^4+{f_7^W} m^2 \nn\\
 &&  +4 {f_8^Z}
   m^4-2 {f_9^Z} m^4+{f_5^Z}
   m^2+{f_6^Z} m^2+{f_7^Z} m^2
\eea
\bea
W_2(E,m,\theta)={f_3^h} m^4-{f_4^h}
   m^4+{f_8^W} m^2-{f_9^W}
   m^2+{f_8^Z} m^2-{f_9^Z} m^2
\eea
where the analytical expressions for the form factors $f_i^{W,Z,h}$ 
is given in \cite{Coriano:2012cr}. In the above expressions we omitted 
the $q^2$ dependence in the form factors $f^{W,Z,h}_k$.

\subsection{Flavour-changing scattering cross section}

In this case the scattering amplitude on an external gravitational field 
$h_{\mu\nu}$,  for on-shell fermions, can be written as
\bea
i\mathcal{S}_{i f}&=& -i\frac{\kappa}{2}\mathcal{N}_i\mathcal{N}_f h_{\mu\nu}\,\hat{T}^{ND \mu\nu }\nn\\
&=&i \frac{G_F}{16 \pi^2 \sqrt{2}}\frac{\kappa}{2}\mathcal{N}_i\mathcal{N}_f h_{\mu\nu}\times\sum_{k=1}^{12} f^{ND}_k(q^2) \hat{O}^{\mu\nu}_k 
\eea
where, for convenience, we have introduced the notation akin to Eq.~(\ref{akin}) 
\beq
\hat{O}_k\equiv \bar u(p_2) O^{\mu\nu}_k u(p_1).
\eeq
The exact analytical expressions for the $f_k(q^2)$ at one-loop, can be found
in Ref.\cite{Coriano:2013msa}.

Then, the flavour-changing changing cross section takes the form  
\beq
\frac{d\sigma}{d\Omega}=\left( \frac{G_F}{16 \pi^2 \sqrt{2}}\right)^2\frac{\kappa^2m_1 m_2}{ 32 \pi^2 E^2}\sqrt{\frac{E^2-m_2^2}{E^2-m_1^2}}\, h_0(q)^2 \,
\mathcal{Y}_w(m_1,m_2),
\eeq
\color{black}
where 
\bea
\mathcal{Y}_w(m_1,m_2)&\equiv& \sum_{i,j=1}^{12}\langle O_i^{} O_j^{  \dagger} \rangle f^{ND}_i (f^{ND}_j)^\dagger \nn\\
&=&  \frac{1}{4 m_1 m_2}Tr\left[ (\slashed p_2 + m_2)  O_i^{ \mu\nu} (\slashed p_1 + m_1) O_j^{\alpha\beta \dagger}\right]\bar{S}_{\mu\nu} \bar{S}_{\alpha\beta}. 
\eea
The expression of $\mathcal{Y}_w$, for general values of the mass parameters, is rather involved. In the case of equal external fermion 
masses $(m_1=m_2)$ we obtain 
\bea
\mathcal{Y}_w(m) &=& \sum_{i=1}^{12}c_i |f^{ND}_i|^2 +2 \sum_{i=1}^{11}\sum_{ j=i+1}^{12}\mathcal{R}e\left( {f^{ND}_i (f^{ND }_j})^\dagger\right)d_{i\,\, j},
\label{non1}
\eea
with the nonvanishing coefficients being given in Appendix \ref{cg}. The cross section above simplifies considerably in the case of massless external fermions, taking the simplified form
\beq
\label{flav}
\frac{d \sigma}{d\Omega}=\left(\frac{G^2_F\kappa^2}{32 \pi^6}\right) h_0^2(q)|f^{ND}_1(q^2)|^2 \cos^2\frac{\theta}{2}, 
\eeq
which involves a single form factor, $f^{ND}_1(q^2)$. We will come back, in a following section, to investigate the behaviour of this cross section at low momentum transfers. \\ From (\ref{flav}) one can consider the possibility of generating flavour-changing transitions between massless external leptons in a gravitational background, which in the neutrino case would 
appear as a gravitational contribution to their flavour oscillations. Implicit in (\ref{flav}) is the presence of a complex factor proportional to the CKM matrix element $V_{\bf i n}$, that we have included into the structure of $f^{ND}_1$.
This factor mediates transitions between a massless external fermion ({\bf i}) and  massive internal ones ({\bf n}), which we sum over. Unitarity of $V$ will rule out such transitions only for mass degenerate internal fermions.

\section{Electromagnetic form factors at small momentum transfers and the cross section} 
In this and in the following section we will expand the form factors at low momentum transfers, starting from the electromagnetic case and then moving to the weak case. This will simplify considerably the expressions of the corresponding cross sections, which 
take, in this limit, rather simple forms.  We start our analysis by observing that the existence of a Ward identity for the conservation of the energy momentum tensor of the various (QED, QCD and electroweak) sectors of the Standard Model has important implications on the structure of most of the form factors which appear in the operatorial expansions presented before. 
Consider the renormalized (with self-energies on the external lines) one-loop correction only, $\hat T^{\mu\nu}$, with on-shell fermions. In this case the EMT Ward identity is given by
\bea
\label{WI2}
q_{\mu} \, \hat T^{\mu\nu}(p_1, p_2)  = 0 \,.
\eea
This describes the conservation of the EMT and emerges from the requirement of invariance under general coordinate transformations. 
For this reason one can reasonably expect that some consequences of the equivalence principle can be extracted from it.
Indeed one can differentiate Eq.~(\ref{WI2}) with respect $q_\alpha$
\bea
\label{DiffWI1}
0 = \frac{\partial}{\partial q_{\alpha}} \bigg[ q_{\mu} \, \hat T^{\mu\nu}(p_1, p_2)  \bigg] =  \hat T^{\alpha\nu}(p_1, p_2) + q_{\mu} \frac{\partial}{\partial q_{\alpha}} \hat T^{\mu\nu}(p_1, p_2) \,,
\eea
and then perform the zero momentum transfer limit setting $q_{\mu} = 0$, ($p_{1 \, \mu} = p_{2 \, \mu}$). The last term in Eq.~(\ref{DiffWI1}) vanishes and we have
\bea
\hat T^{\mu\nu}(p_1, -p_1) \equiv  \hat T^{\mu\nu}(p, q=0) = 0 \,.
\eea
We then conclude that the Ward identity implies the vanishing of the one-loop corrected $\hat T^{\mu\nu}(p_1,p_2)$ matrix element for on-shell fermions in the zero momentum transfer limit. \\
This has some interesting consequences on the form factors. In this respect one has to take into account, at the same time, the tensor expansion of the correlator. Notice that in the $q_\mu = 0$ limit, the tensor structures $O_{V 4}$, $O_{C  5}$, $O_{C  6}$ go to zero and nothing can be said on the corresponding form factors. But for the remaining elements of the basis, which survive for $q_\mu = 0$, the corresponding form factors must vanish for $q^2= 0$.

The simplest case in which one can perform the small $q^2=-Q^2$ limit of the form factors is the electromagnetic one. 
In this case we expand in powers of $Q^2/m^2$ and obtain the expressions up to $ O((Q^2)^{3/2})$
\bea
f_1&=&-\frac{\pi^2\sqrt{Q^2}}{8 m} + \frac{Q^2}{72 \,m^2}\left( 9 \log\frac{m^2}{Q^2} + 12 \log\frac{\lambda^2}{Q^2} + 65 - 36 \log 2\right) \nn\\
f_2&=&\frac{\pi^2\sqrt{Q^2}}{4 m} +\frac{Q^2}{36 m^2}\left( -40 + 36 \log 2 + 33\log\frac{Q^2}{m^2}\right) \nn\\
f_3&=& -\frac{\pi^2\sqrt{Q^2}}{16 m^3} +\frac{Q^2}{144 m^4} 
\left( 2 - 36 \log 2 -57 \log\frac{Q^2}{m^2}\right) \nn\\
f_4 &=&\frac{\pi^2}{4 m \sqrt{Q^2}}+ 
\frac{1}{36 m^2}\left( -40 + 36 \log 2 + 33\log\frac{Q^2}{m^2}\right) \nn\\
\label{fEM}
\eea
The values at the leading order expansion in $q^2$ of the form factors in Eq.~(\ref{fEM}) are in agreement with the corresponding results of Ref.\cite{Berends:1975ah}.

Notice the presence of an infrared cutoff $\lambda$ in the form factors
in Eq.~(\ref{fEM}), where $\lambda$ is the photon mass used to regularize the infrared divergences.
This is due to the exchange of a massless photon, generated by the 3-point scalar integral expanded around zero $Q^2$
\bea
\mathcal C_0 (Q^2=0, m^2, m^2) = \frac{x_s}{m^2 (1 - x_s^2)} \bigg\{  - 2 \log \frac{\lambda}{m}  \log x_s + \ldots \bigg\}
\eea 
where the ellipsis mark some infrared finite terms. We have defined
\beq
x_s = - \frac{1-\beta}{1+\beta} \qquad \beta = \sqrt{1 - 4 m^2 / q^2}. 
\eeq
It can be shown by a direct computation in QED, that the infrared singular parts of the virtual corrections of the matrix element $G\to f\bar{f}$, with 
$G$ being the graviton, once summed, are proportional to the tree-level vertex. A second infrared singular contribution comes from the renormalization counterterm of the masless photon, which alse needs to be taken into account. As shown in \cite{Coriano:2012cr}, the inclusion of the real photon emissions from the fermion legs in the collinear limit (the process $G\to f\bar{f} +\gamma$), regulated by a cutoff $\lambda$ and integrated over the photon, allows to remove the singularity in $\log(\lambda)$. 
  
Using these expressions, the virtual plus soft corrections to the cross section - at leading order in $Q^2/m^2$ - take the form
 \beq
\frac{d \sigma}{d \Omega}=\frac{(G M)^2}{\sin^4\frac{\theta}{2}}\left(F^{(0)}(E,m,\theta) + \frac{\alpha e_Q^2}{4 \pi} 
\tilde{F}_D^{(1)}(E,m,\theta)  \right),
 \eeq
where $F^{(0)}$ is given by (\ref{F0}) and 
\bea
\tilde{F}^{(1)}_D(E,m,\theta) &=& a_1 \sin\left(\frac{\theta}{2}\right) + 
a_2\sin^3\left(\frac{\theta }{2}\right) + a_3\sin^5\left(\frac{\theta }{2}\right)\nn
\eea
with 
\bea
a_1=-\frac{\pi^2}{2 m}\frac{6 E^4 - 5 E^2 m^2 + m^4}{ |\vec{p}_1|^3} \qquad a_2=-\frac{3\pi^2}{4 m} 
\frac{-4 E^2 + m^2}{|\vec{p}_1|}\qquad a_3=-\frac{\pi^2}{4 m} |\vec{p}_1|.
\eea

\section{Weak form factors at small momentum transfers}
Now, we focus on the kinematical region where $q^2 \ll m_{EW}^2$
and perform an expansion of the $\hat T^{\mu\nu}_Z(p_1,p_2)$ in powers of $q^2/m_{EW}^2$, where
$q_{\mu}=p_1-p_2$ is the transfered momentum and
$m_{EW}$ stands for any characteristic electroweak scale, identified with the
$W,Z$ or Higgs boson mass.
Because of the conservation of the matter energy momentum tensor  $T_{\mu\nu}$
on shell (which holds only in the absence of gravitational field corrections), it can be easily show that 
all the form factors
$f^{Z,W,h}_k(q^2)$ in Eq.~(\ref{T}), except for  $f^{Z,W,h}_5(q^2)$,
vanish in the limit $q^2\to 0$ and turn out to be of order 
${\cal O}(q^2)$. We will start first our discussion with the flavour-conserving
transitions, and then we will address the flavour-changing ones.

\subsection{  Flavour-diagonal transitions}
Here we provide the expressions for $\hat T^{\mu\nu}(p_1,p_2)$ at the
leading  term in the $q^2G_F$ and external masses $m^2G_F$ expansion. 
In the case of
 $Z$ and $W$ exchanges, it is
convenient to express  Eq.~(\ref{T}) in terms of rescaled operators as
\bea
\hat T^{\mu\nu}_{Z,W}(p_1,p_2) &=& i \, \frac{G_Fq^2}{16 \pi^2 \sqrt{2}}   \sum_{k=1}^6  \bar{f}^{W,Z}_k \, \bar u(p_2) \, \bar{O}^{\mu\nu}_{C  k} \, u(p_1) \,, 
\label{T}
\eea
where
$\bar{O}^{\mu\nu}_{C  i} = O^{\mu\nu}_{C  i}$ for $i=1,2,3$ and
\beq
\bar{O}^{\mu\nu}_{C  4} = \frac{1}{m_{W,Z}^2}O^{\mu\nu}_{C  4},
 \qquad
\bar{O}^{\mu\nu}_{C  5} = \frac{1}{q^2}O^{\mu\nu}_{C  5},
 \qquad
\bar{O}^{\mu\nu}_{C  6} = \frac{1}{q^2}O^{\mu\nu}_{C  6}\, ,
\label{chiralbasis2}
\eeq
such that all the form factors  $\bar{f}^{W,Z}_k$ are now dimensionless.

Analogously, for the Higgs boson corrections, we define the rescaled operators
\bea
\hat T^{\mu\nu}_{h}(p_1,p_2) &=& i \, \frac{G_Fq^2}{16 \pi^2 \sqrt{2}}   \sum_{k=1}^4  \bar{f}^{h}_k \, \bar u(p_2) \, \bar{O}^{\mu\nu}_{V  k} \, u(p_1) \,, 
\label{T}
\eea
where
$\bar{O}^{\mu\nu}_{V  i} = O^{\mu\nu}_{V  i}$ for $i=1,2$, and
\beq
\bar{O}^{\mu\nu}_{V  3} = \frac{1}{m_{h}^2}O^{\mu\nu}_{V  3}, \qquad
\bar{O}^{\mu\nu}_{V  4} = \frac{1}{q^2}O^{\mu\nu}_{V  4}\, ,
\eeq
with all form factors  $\bar{f}^{h}_k$ dimensionless.
In the case of the $Z$ and Higgs boson corrections, we will provide the
results for the form factors by retaining the external fermion-mass dependence up to the second order ${\cal O}(x_{Z,h}^2)$, with $x_{Z,h}=m^2/m_{Z,h}^2$. In the case of $W$ corrections, we will give only the results at the leading order in the external masses, while retaining the exact analytical dependence in the internal loop masses.

Then, for the case of the $W$ and $Z$ corrections we obtain 

\bea
\bar{f}^{Z}_1 &=& 
\left(1+8 a v+4 v^2\right)\left(\frac{11}{72} 
+\frac{x_Z}{24}\log (x_Z)\right)
  + \frac{x_Z}{288} \left(17 - 40
   a v-60 v^2\right)
\nonumber
\\
\bar{f}^{Z}_2 &=&
\left(1-8 a v+4 v^2\right)\left(\frac{11}{72} 
+\frac{x_Z}{24}\log (x_Z)\right)
  + \frac{x_Z}{288} \left(17 +40
   a v-60 v^2\right)
\nonumber
\\
\bar{f}^{Z}_3 &=&
\frac{1}{6} \left(\left(2-8 v^2\right) \log (x_Z)+1-20
   v^2\right)+\frac{x_Z}{24}  \left(4 \left(7-44
   v^2\right) \log (x_Z)+27-260
   v^2\right)
\nonumber
\\
\bar{f}^{Z}_4 &=&
\frac{3-4 v^2}{72}+\frac{x_Z }{360}\left(120
   \left(1-2 v^2\right) \log (x_Z)+301-716
   v^2\right)
\nonumber
\\
\bar{f}^{Z}_5 &=& -\bar{f}^Z_3
\nonumber
\\
\bar{f}^{Z}_6 &=& \bar{f}^Z_2-\bar{f}^Z_1\, ;
\eea

\bea
\bar{f}^{W}_{i} &=&\sum_f \lambda_f g_{i}(x_f) 
\eea
for $i=1,2,3,4$ while, due to the Ward identity,
\beq
\bar{f}^{W}_5 = -\bar{f}^W_3,\qquad
\bar{f}^{W}_6 = \bar{f}^W_2-\bar{f}^W_1,\, 
\eeq
and the $g_i(x)$ functions are given by
\beq
g_1(x)=g_a(x),
\qquad
g_2(x)=x_W g_d(x),
\qquad
g_3(x)=-g_b(x),
\qquad
g_4(x)=g_c(x),
\eeq
while, due to the Ward identities, we have
\beq
\bar{f}^{W}_5 = -\bar{f}^W_3
\qquad \bar{f}^{W}_6 = \bar{f}^W_2-\bar{f}^W_1.\, 
\eeq
The explicit expressions of the $g$ functions can be found in Appendix \ref{cg}.
 In the case of the form factors $\bar{f}^h$ involving a virtual Higgs boson 
we obtain 
\bea
\bar{f}^{h}_{1} &=&\frac{x_h}{18}
\nonumber \\
\bar{f}^{h}_{2} &=& x_h \left(\frac{\log (x_h)}{3}+\frac{11}{6}-6\chi\right)
\nonumber \\
\bar{f}^{h}_{3} &=&\frac{x_h}{6} \left(3+\log (x_h)\right)\, ,
\eea
where, due to the Ward identities,
\bea
\bar{f}^h_{4} &=&-\bar{f}^h_2\, ,
\eea
with
$x_f=m_f^2/m_W^2$ and  $\lambda_f=V_{\bf if}V^{\star}_{\bf if}$, where $i$ and
$f$ indicate the external and internal flavour indices respectively, while
$V_{\bf if}$ stands for the corresponding CKM matrix element.

\subsection{Flavour-changing transitions}
Now, we perform the expansion of the twelve form factors $f_k^{ND}$
appearing in Eq.~(\ref{TFC}) at the first order in $q^2G_F$, as in the 
flavour-conserving transitions, by retaining only the leading contribution in the external masses. Due to the fact that in the flavour-changing case 
$m_1\neq m_2$, some of the form factors will not vanish in the $q^2\to 0$ 
limit. Therefore, it is convenient to keep the notation of form factors 
$f_k^{ND}$ as in the basis of Eq.~(\ref{TFC}), where they have different dimensions.

Finally, the non-vanishing expressions of the form factors, up to the order 
$q^2$, where we neglect terms of order ${\cal O}(q^2 m_{1,2}^2/m_W^4)$ and 
${\cal O}(m_{1,2}^4/m_W^4)$, are given by

\beq
\begin{array}{ll}
f_1(q^2)=q^2\sum_f \lambda_f  g_a(x_f)  &  f_2(q^2)=-\left(m_1^2-m_2^2\right)\sum_f \lambda_f g_a(x_f)  \\
f_3(q^2)=- q^2 \sum_f \lambda_f g_b(x_f) & f_4(q^2)=
\frac{q^2}{m_W^2}\sum_f \lambda_f g_c(x_f)   \\
f_5(q^2)=
\sum_f \lambda_f \left(g_b(x_f) -\frac{q^2}{m_W^2} g_g(x_f)\right) & f_6(q^2)=-
\frac{\left(m_1^2-m_2^2\right)}{m_W^2} \sum_f \lambda_f g_c(x_f) \\ 
f_7(q^2)=2
\left(m_1^2-m_2^2\right)\sum_f \lambda_f g_a(x_f)&f_{10}(q^2)=-
\sum_f \lambda_f \left(g_a(x_f)
+\frac{q^2}{m_W^2} g_h(x_f)\right) \\
f_{11}(q^2)=q^2\sum_f \lambda_f  g_d(x_f) & f_{12}(q^2)=-
\left(m_1^2-m_2^2\right)\sum_f \lambda_f g_d(x_f)
\end{array} 
\label{FF}
\eeq
where 
$g_W$ is the weak coupling, and $\lambda_f\equiv 
V_{\bf 1f} V^{\star}_{\bf 2f}$, with $V_{\bf ij}$ the corresponding CKM matrix element. 
The expressions for the functions $g_{a-d}(x)$ are reported in Appendix B.
We do not report the expression of the form factors $f_{8,9}$ 
since they are vanishing at the leading order in our approximation.
The unitarity of the CKM is expressed by the relation $\sum_f \lambda_f=0$.
One can easily check that in the adopted approximation,
the form factors in Eq.~(\ref{FF}) satisfy the Ward identity reported 
in \cite{Coriano:2012cr}.

\section{The effective Hamiltonians and the cross sections at low momentum transfers: massive gravity}
In this section we turn to the computation of the weak corrections to the diagonal and non-diagonal cross sections in the limit of small momentum transfers. At the same time we will derive the expression of the effective Hamiltonian describing the transition from the initial to the final state in the presence of an external gravity source. In particular, we will include in our analysis also the case in which the graviton has a small mass $m_G$, detailing the modifications on the cross sections and on the Hamiltonian respect to the case of General Relativity (GR). The difference between the massive and the massless graviton cases can be worked out more directly 
starting from the standard scattering amplitude expressed in terms of the Feynman propagator. Here we discuss the amplitude in this form, 
performing the heavy source limit at the final stage. 

We recall that in momentum space, the  2-particle covariant gravitational scattering amplitude amplitude is
given by
\bea
{\cal M}=8\pi G\, \left(\hat{T}_{\mu\nu}~ 
P_h^{\mu\nu\alpha\beta}(q^2) ~ \hat{T}^{\rm ext}_{\alpha\beta}\right),
\label{Mamp}
\eea
where $P_h^{\mu\nu\alpha\beta}(q^2)$ is the graviton propagator in momentum
space, and where $q=p_1-p_2$, 
and $\hat{T}^{\rm ext}_{\mu\nu}$ is the Fourier transform of the EMT associated to the external source.

In the Einstein theory, 
the graviton propagator in a covariant gauge is given by
\bea
P_h^{\mu\nu\alpha\beta}(q^2)=\frac{i}{q^2-i\varepsilon}\frac{1}{2}\left(
\eta^{\mu\alpha}\eta^{\nu\beta} + 
\eta^{\nu\alpha}\eta^{\mu\beta} - \eta^{\mu\nu}\eta^{\alpha\beta}
+ Q^{\mu\nu\alpha\beta}(q)\right),
\label{EinstProp}
\eea
where the last term  $Q^{\mu\nu\alpha\beta}(q)$, which is gauge dependent, 
is a tensor made by linear combinations
of an even number of $q$ momenta with open indices, 
such as for instance $q^{\mu}q^{\nu} \eta^{\alpha\beta}$ or
$q^{\mu}q^{\nu} q^{\alpha}q^{\beta}$.
Nevertheless, due to  the conservation of the
energy-momentum tensor, the 
contribution of $Q^{\mu\nu\alpha\beta}(q)$ vanishes when contracted with 
$\hat{T}_{\mu\nu}$ or $\hat{T}^{\rm ext}_{\alpha\beta}$, leading
to a gauge-invariant result.
In a scenario in which the graviton has a small mass 
$m_G$, the corresponding graviton propagator in the unitary gauge takes a modified form 
\bea
P_{G}^{\mu\nu\alpha\beta}(q^2)= \frac{i}{q^2-m_G^2-i\varepsilon}
\frac{1}{2}\left(
\tilde{\eta}^{\mu\alpha}\tilde{\eta}^{\nu\beta} + 
\tilde{\eta}^{\nu\alpha}\tilde{\eta}^{\mu\beta} - 
\frac{2}{3}\tilde{\eta}^{\mu\nu}\tilde{\eta}^{\alpha\beta}
\right),
\label{PropGm}
\eea
where $\tilde{\eta}^{\mu\nu} \equiv \eta^{\mu\nu} - q^\mu q^\nu/m_G^2$.
As one can see, there is a discontinuity in the above propagator in the massless 
graviton limit, which differs by a finite term from the massless case.

The scattering amplitude in the weak field approximation, in General Relativy (GR) is given by
\bea
{\cal M}_{GR}&=&
\frac{i 4\pi G}{\left(q^2-i\varepsilon\right)}
\left[2
\hat{T}^{\mu\nu}\, \hat{T}^{\rm ext}_{\mu\nu}-
\, \hat{T}^{\mu}_{\mu} \hat{T}^{\rm ext~ \nu}_{\nu}\right],\, 
\label{Mc}
\eea
while for a massive graviton (MG) it is given by
\bea
{\cal M}_{MG}&=&
\frac{i4\pi G}{\left(q^2-m_G^2-i\varepsilon \right)}
\left[2
\hat{T}^{\mu\nu}\, \hat{T}^{\rm ext}_{\mu\nu}-
\, \frac{2}{3}\hat{T}^{\mu}_{\mu} \hat{T}^{\rm ext~ \nu}_{\nu}\right]\, .
\label{Mc}
\eea
At this point it is convenient to decompose the amplitude as
\bea
{\cal M} &=& {\cal M}^{\rm tree} + \delta {\cal M},
\eea
where  ${\cal M}^{\rm tree}$ stands for the usual tree-level contribution
and $\delta {\cal M}$ contains the one-loop EW corrections to the EMT.
Now we will analyze the two cases of flavour-conserving and flavour-changing
transitions.

\subsection{Flavour-conserving transitions}
By applying the on-shell relations for the external fermion states and the 
condition that the external source is conserved, we obtain for the 
EW one-loop contributions $\delta {\cal M}=\delta {\cal M}^{W}+\delta {\cal M}^{Z}+\delta {\cal M}^{h}$ we obtain 
\bea
\delta {\cal M}^{W,Z}&=&\frac{1}{\Lambda^4_{\rm eff}}\sum_{k=1}^4
\Big\{2 \bar{f}^{W,Z}_k \, \bar u(p_2) \, \bar{O}^{\mu\nu}_{C  k} \, u(p_1) \, \hat{T}_{\mu\nu}^{\rm ext}
-C\bar{f}^{W,Z}_k \, \bar u(p_2) \, \bar{O}^{\mu}_{\mu\, C  k} \, u(p_1)\, \hat{T}_{\nu}^{\rm ext ~ \nu }\Big\}
\label{HeffWZ}
\eea
for the corrections involving the exchanges of $W$ and $Z$ gauge bosons in the loop and 

\bea
\delta {\cal M}^{h}&=&\frac{1}{\Lambda^4_{\rm eff}}\sum_{k=1}^3
\Big\{2 \bar{f}^{h}_k \, \bar u(p_2) \, \bar{O}^{\mu\nu}_{V  k} \, u(p_1) \, \hat{T}_{\mu\nu}^{\rm ext}
-C\bar{f}^{h}_k \, \bar u(p_2) \, \bar{O}^{\mu}_{\mu\, V  k} \, u(p_1)\, \hat{T}_{\nu}^{\rm ext ~ \nu }\Big\}
\label{HeffH}
\eea
for those related to the exchanges of a Higgs field. 
The parameter $C$ allows to describe both the massless and the massive gravity scenarios, with $C=1$ and $C=2/3$ respectively, while
the effective scale $\Lambda_{\rm eff}$ is given by
\bea
\Lambda_{\rm eff}&=&\left(\frac{2\pi\sqrt{2}}{G G_F}\right)^{1/4} \simeq~ 10^{11}\,  {\rm GeV}\, .
\eea
As we can see, the characteristic energy scale $\Lambda_{\rm eff}$ of the 
local interactions
is almost 8 order of magnitudes smaller than the Planck scale. In the case of massive gravity scenarios
we have assumed that $q^2 \gg m_G^2$ and neglected terms of order 
${\cal O}(m_G^2/q^2)$ in the graviton propagator.
These results show that the contribution of the 
EW corrections to the flavour-diagonal (FD) gravitational scattering of 
fermions on an external gravitational source, in the kinematic region $|q^2| \ll m_W^2$, can be described by a local effective Hamiltonian given by
\bea
{\cal H}^{\scriptscriptstyle {\rm FD}}_{\rm eff}&=&\frac{1}{\Lambda_{\rm eff}^4} \left(\sum_{k=1}^4\bar{f}^{W,Z}_k Q_{C_k}
+\sum_{k=1}^3\bar{f}^{h}_k Q_{V_k}\right),
\eea
where $Q_{C_k}$ and $Q_{V_k}$ are local 8-dimensional operators for the 
$W,Z$ and Higgs boson corrections respectively, defined as
\beq
Q_{C_k}\sim [\bar{\psi}(x)\, \bar{O}^{\mu\nu}_{C  k}\, \psi(x)] T_{\mu\nu}^{\rm ext}(x), 
\qquad Q_{V_k}\sim [\bar{\psi}(x)\, \bar{O}^{\mu\nu}_{V  k}\, \psi(x)] T_{\mu\nu}^{\rm ext}(x),
\eeq
with $\psi(x)$ standing for a generic SM fermionic field.
The characteristic energy scale  $\Lambda_{\rm eff}^4$ is defined above.
The tree-level matrix elements of the effective Hamiltonians given above, taken between on-shell external fermion states, match the results obtained in Eqs.(\ref{HeffWZ}) and (\ref{HeffH}) by using the full Standard Model theory.

\subsection{Flavour-changing transitions}
Following the case of the flavour-conserving transitions,  we consider now the 
flavour-changing scattering amplitude on an external gravitational source
\bea
f_1(p_1) + T^{\rm ext} \to f_2(p_2) + T^{\rm ext}\, ,
\eea
where we will assume the two fermions $f_1$ and $f_2$ having different flavour 
and masses $m_1\neq m_2$.
Despite the presence of the $1/q^2$ pole in the graviton propagator, 
the flavour-changing gravitational scattering in General Relativity turns out to be local.
Indeed, we find
\bea
{\cal M}^{\rm FC}_{m_G=0} &=& \frac{1}{\Lambda_{\rm eff}^4} \sum_f \lambda_f
\bar{u}(p_2)\left\{ \Big(\gamma^{\mu} p^{\nu}+\gamma^{\nu} p^{\mu}\Big)P_L 
g_a(x_f) - g^{\mu\nu}M_{+}\Big(C\,g_a(x_f)-(\frac{3}2\,C -1 ) g_b(x_f)\Big)
\right. \nonumber \\
&+& \left. 
g^{\mu\nu} M_{-} \frac{(p\cdot q)}{(q^2-i\varepsilon)} g_a(x_f) 
\left[2+C\left(- g^{\alpha}_{\alpha} +2\right)\right]
\right\} u(p_1)\, 
\hat{T}^{\rm ext}_{\mu\nu}
\label{Mgeneral}~.
\eea
The second line in Eq.~(\ref{Mgeneral}) is zero if C=1 (massless graviton) and 
$g^{\alpha}_{\alpha}=4$, showing that in the Einstein theory of 
General Relativity the $1/q^2$ pole  cancels out.
This is a general result that holds also in the exact case, as can
be easily proved  using the Ward identities. 
As we can see from Eq.~(\ref{Mgeneral}), the flavour-changing gravitational
scattering on an external source - for the $\Delta F=1$ flavour 
transitions - can be described by a local effective Hamiltonian given by
\bea
{\cal H}^{\rm FC}_{\rm eff}|_{m_G=0}=\frac{1}{\Lambda_{\rm eff}^4} 
\left( Q_{\rm \scriptscriptstyle{FC} 1}^{\mu\nu}\sum_f \lambda_f g_a(x_f) +
Q_{\rm \scriptscriptstyle{FC} 3}\sum_f \lambda_f( g_b(x_f)-g_a(x_f))  \right),
\label{HeffG}
\eea
where the $Q_{\rm \scriptscriptstyle{FC} 1,3}$ are local eight-dimensional flavour-changing 
operators, which are defined as 
\beq
Q_{\rm \scriptscriptstyle{FC} 1,3}= [\bar{\psi}_2(x)\, O^{\mu\nu}_{1,3}\, \psi_1(x)] T_{\mu\nu}^{\rm ext}(x),
\eeq
 with the basis $O^{\mu\nu}_{1,3}$ defined in Eq.~(\ref{basis}). In the expression above, 
$\psi_{1}$ and $\psi_{2}$, stand for the fermion fields corresponding to  the initial and final fermions with flavour 1 and 2 respectively, similarly to the diagonal case, but now of different flavours.

The amplitude for the flavour-changing gravitational scattering, in the case in which the graviton has a small but non-vanishing mass,
has a remarkable difference with respect to the massless graviton case. Due to the different structure of the massive graviton propagator (C=2/3), we obtain
\bea
{\cal M}^{\rm FC}_{m_G\neq 0} &=& \frac{1}{\Lambda_{\rm eff}^4} \sum_f \lambda_f
\bar{u}(p_2)\left\{ \frac{q^2}{(q^2-m_G^2-i\varepsilon)}
\left( \Big(\gamma^{\mu} p^{\nu}+\gamma^{\nu} p^{\mu}\Big)P_L 
 - \frac23 g^{\mu\nu}M_{+}\right) g_a(x_f)
\right. \nonumber \\
&+& \left.
\frac23 g^{\mu\nu} M_{-} \frac{(p\cdot q)}{(q^2-m_G^2-i\varepsilon)} g_a(x_f)
\right\} u(p_1)\, 
\hat{T}^{\rm ext}_{\mu\nu}
\label{Mgmass}~.
\eea
As we can see from  Eq.~(\ref{Mgmass}), the
$1/q^2$ pole in the last term of the amplitude does not vanish
in the limit $m_G\to 0$. The residue of the pole 
is proportional to $(p\cdot q)$, 
which in the case of on-shell external fermions is equal to 
$(p\cdot q)=m_1^2-m_2^2$. This has important implications in the case of a flavour-changing scattering for massive gravity or in the interaction with a graviscalar potential. Indeed,  the last term in Eq.~(\ref{Mgmass}) 
generates an off-diagonal contribution (in flavour space) to the Newton
potential, inducing long-distance flavour-changing interactions.
On the other hand, the local part of the interaction can be described by the following effective Hamiltonian which is similar to the one in Eq.~(\ref{HeffG}), namely
\bea
{\cal H}^{\rm FC}_{\rm eff}|_{m_G\neq 0}=\frac{1}{\Lambda_{\rm eff}^4} 
\left( Q_{\rm \scriptscriptstyle{FC} 1}^{\mu\nu}\sum_f \lambda_f g_a(x_f) -\frac{2}{3}
Q_{\rm \scriptscriptstyle{FC}3}^{\mu\nu}\sum_f \lambda_f g_a(x_f)  \right)\, ,
\label{HeffG}
\eea
where we have assumed the graviton mass to be $m_G^2\ll |q^2|$
and neglected terms of order  ${\cal O}(m_G^2/q^2)$ in the graviton propagator.

\subsection{Flavour-changing scattering cross sections at low momentum transfers}
The behaviour of the cross sections at low momentum transfers is particularly interesting in the flavour-changing case, 
due to the possibility of generating transitions in the presence of a gravitational source. As such they may appear as additional contributions, though largely suppressed, to ordinary flavour oscillations. We will start by considering the case of massive gravity, where this phenomenon has been pointed out before \cite{Degrassi:2008mw}. 
\begin{itemize} 
\item{\bf Massive gravity scenario}\\
We consider the anelastic flavour-changing scattering of fermions
on an external gravitational source in the framework of a massive gravity scenario. Generalization to the case of graviscalar interactions are straightforward. In particular, we consider the scattering on a heavy point-like particle $T$ 
\bea
f_1(p_1) + T \to f_2(p_2) + T,
\label{MFC}
\eea
where we assume that the particle T is very heavy and acts as a generator of a static gravitational field. The fermions $f_{1,2}$ have masses $m_{1,2}$ respectively. We assume that the gravitational field has a small mass in order to generate a long distance potential.
The amplitude of the process can be taken from Eq.~(\ref{Mgmass}) by retaining
only the last term in parenthesis, proportional to the $1/(q^2-m_G^2)$ pole, 
and neglecting terms of order ${\cal O}(q^2)$.
Then, 
the leading contribution to the 
amplitude for the process in (\ref{MFC}) reduces to 
\bea
{\cal M}_{12}&\simeq & \frac{1}{\Lambda_{\rm eff}^4} \sum_f \lambda_f
\bar{u}(p_2)\left\{ \frac{2}{3}
M_{-} \frac{(p\cdot q)}{(q^2-m_G^2-i\varepsilon)} g_a(x_f)
\right\} u(p_1)\, 
\hat{T}^{{\rm ext}~\mu}_{\mu},
\eea
where $\hat{T}^{{\rm ext}~\mu}_{\mu}= g^{\mu\nu}\hat{T}^{\rm ext}_{\mu\nu}$, and 
$\hat{T}^{\rm ext}_{\mu\nu}$ stands for the Fourier transform of the energy momentum tensor associated to the heavy external particle $T$.
If we consider the external source $\hat{T}^{\rm ext}_{\mu\nu}$ to be a point-like heavy source as in (\ref{emt}), we then have
\bea
{\cal M}_{12} &=& \frac{M}{\Lambda_{\rm eff}^4} \sum_f \frac{2}{3}\lambda_f g_a(x_f)
\left\{\bar{u}(p_2)
 M_{-} \frac{(p\cdot q)}{(q^2-m_G^2-i\varepsilon)}
u(p_1)\right\},\, 
\label{M12}
\eea
Now, we neglect the contribution of the $m_G$ in the denominator
of Eq.~(\ref{M12}), since we assume that $m_G^2 \ll |\vec{q}|^2$.
By summing over the polarizations and mediated over the initial ones, we get
\bea
\frac{d \sigma_{12}}{d \Omega} &=& 
\frac{(1+\Delta)}{16 \pi^2}(G_N M)^2 G_F^2(m_1^4-m_2^4)|C|^2
\frac{\Delta\left(1-\rho\cos{\theta}+\frac{\Delta m_1^2}{m_1^2+m_2^2}\right)}{
8\pi^2\left(2\left(1-\rho \cos{\theta}\right)+\Delta\right)^2}
\label{M12Q}
\eea

where
\bea
\rho=\sqrt{1+\Delta}\, ,  ~~~~~ \Delta=\frac{m_1^2-m_2^2}{p^2}, ~~~ 
C=\frac{2}{3}\sum_f \lambda_f g_a(x_f)\, .
\eea
and $p$ indicates the modulus of the incoming 3-momentum, namely
$p \equiv |\vec{p}_1|=\sqrt{E^2-m_1^2}$, with $E$ the energy of the incoming
particle.

Now from Eq.~(\ref{M12Q}) we can see that the angular distribution is fully 
peaked  in the forward direction $\theta\simeq 0$, and there are no infrared divergences in the $\theta=0$ limit due to the fact that $m_1\neq m_2$. Notice that the term $\Delta$ plays the role of a small parameter as the term $\alpha$ does in the elastic case. 

Finally, by integrating over the scattering angle, the total cross section takes the form
\bea
\sigma_{12}&=&
\sigma_{12}^0
\Big\{
1+\frac{(m_1^2+m_2^2)\Delta}{4(m_1^2-m_2^2) \rho}\log{
\left(\frac{2+\Delta+2\rho}{2+\Delta-2\rho}\right)}
\Big\}(1+\Delta),
\label{sigmaFC}
\eea
where $\sigma_{12}^0$ is the leading term 
in the zero order expansion in $\Delta$. This is given by
\bea
\sigma_{12}^0&=&
\frac{(G_N M)^2 G_F^2 \left(m_1^2-m_2^2\right)^2 |C|^2}{64 \pi^3},
\label{sigma0}
\eea
which depends only on the mass splittings and on other fundamental constants, but not on the energy. We stress that the 
above results are obtained in the approximation of small momentum transfers, $|q^2| \ll m^2_W$, hence the
result in Eq.~(\ref{sigmaFC}) holds only in the low energy regime $E \ll m_W$.

\item{\bf Massless gravity}\\
\end{itemize}
In the case of ordinary gravity, it is interesting to consider massless external fermions. The flavour-changing cross section, whose general expression has been given in Eq.~(\ref{flav}), in this limit takes the simplified form 

\beq
\label{flav1}
\frac{d \sigma}{d\Omega}=\left( \frac{G_F}{\sqrt{2}}\right)^2 \frac{(2 G M)^2}{\pi^4}|\sum_f \lambda_f g_a(x_f)|^2 \, \cos^2{\frac{\theta}{2}}\, .
\eeq
Notice that if all the internal fermions which appear in the loop corrections are mass degenerate - and this 
covers also the complete massless case - then the unitarity of the CKM matrix causes the cross section to vanish. On the other hand,  Eq.~(\ref{flav1}) shows that an external gravitational field may induce, in general, flavour-changing transitions. The result in Eq.~(\ref{flav1}) holds in the range of 
$|q^2| \ll m^2_W$.

\section{Diagonal  weak form factors and cross section at large momentum transfers} 

Before coming to our comments and conclusions, we present the expression of the related cross sections at large momentum transfers, in the cases in which we drop the masses of the external fermions.   
In the non-diagonal case this study, in the weak case, has been presented in \cite{Degrassi:2008mw}, while in the QED case the diagonal cross section at large $q^2$ has been given in \cite{Berends:1975ah}. We consider the limit of large and spacelike momenta $q^2=-\vec{q}^{\,2}$, where we drop all the contributions proportional to the fermion mass $m$. The expansion is organized in terms of $M_Z^2/q^2$, with the only non vanishing form factors being given by
 \bea
 f^Z_1 &=& \frac{1}{2} {m_Z}^2 \left(-5 a^2-(a+v)^2 \log
   \left(-\frac{{m_Z}^2}{\text{qs}}\right) \left(2 \log
   \left(-\frac{{m_Z}^2}{{q^2}}\right)+3\right)-12 a v-5
   v^2\right) \nn\\
   f^Z_2 &=& \frac{1}{2} {m_Z}^2 \left(-5 a^2-(a-v)^2 \log
   \left(-\frac{{m_Z}^2}{{q^2}}\right) \left(2 \log
   \left(-\frac{{m_Z}^2}{{q^2}}\right)+3\right)+12 a v-5
   v^2\right) \nn \\
  f^Z_3 &=&2 {m_Z}^2 \left(\left(v^2-7 a^2\right) \log
   \left(-\frac{{m_Z}^2}{{q^2}}\right)+4 \left(v^2-3
   a^2\right)\right) \nn\\
 f^Z_4&=&  -\frac{2 {m_Z}^2}{{q^2}} \left(4 a^2 \log
   \left(-\frac{{m_Z}^2}{{q^2}}\right) \left(\log
   \left(-\frac{{m_Z}^2}{{q^2}}\right)+4\right)+25
   a^2+v^2\right) \nn\\
\eea
\bea
f^Z_5&=&  -\frac{{m_Z}^2}{{q^2}} \left(2 \left(v^2-7 a^2\right) \log
   \left(-\frac{{m_Z}^2}{{q^2}}\right)-25 a^2+7
   v^2\right) \nn\\
 f^Z_6&=&  \frac{a {m_Z}^2 v}{{q^2}} \left(4 \log
   ^2\left(-\frac{{m_Z}^2}{{q^2}}\right)+6 \log
   \left(-\frac{{m_Z}^2}{{q^2}}\right)+11\right). \nn\\
\eea
A similar limit on the form form factors with a virtual $W$ gives 
  \bea
  f^W_1 &=&\frac{{m_W}^2}{4 ({x_f}-1)^2}\times \nn\\
  && \times \left[ (-1 + x_f) (22 + x_f (-31 + 3 x_f) + 
    2 (-1 + x_f) (-6 + x_f - 2 \log\left(-\frac{m_W^2}{q^2}\right)) \log\left(-\frac{m_W^2}{q^2}\right))\right.  \nn\\
  &&\left. \qquad \qquad + 2 (-4 + x_f) (-2 + x_f) x_f \log x_f - 16 (-1 + x_f)^2 \textrm{Li}_2( 1 - x_f])\right] \nn 
  \eea
  \bea
f^W_2&=&  -\frac{{m_W}^2 ({x_f}+2) \left({x_f}^2-2 {x_f}
   \log ({x_f})-1\right)}{4 ({x_f}-1)^2} \nn \\
f^W_3 &=& \frac{{m_W}^2}{{x_f}-1}\times \nn\\
&&\left[ {x_f} (4 \chi -4 \chi  {x_f}+{x_f}-1) \log ^2\left(-\frac{{m_W}^2}{{q^2}}\right)-8 ({x_f}-1) ((\chi -1)
   {x_f}+1)\right. \nn\\
   && \left. -{x_f} \log ({x_f}) (8 \chi {x_f}+(4 \chi -1) ({x_f}-1) \log ({x_f})-5 {x_f}+6) \right.\nn\\
&&\left. (1- {x_f}) \log \left(-\frac{{m_W}^2}{{q^2}}\right) (8 \chi  {x_f}+2 (4 \chi -1) {x_f} \log ({x_f})-5
   {x_f}+6)\right. \nn\\
  &&\left. -4 (4 \chi -1) ({x_f}-1) {x_f} \textrm{Li}_2\left(1-\frac{{m_W}^2}{{m_f}^2}\right)\right]
\eea  
where we have set $x_f=m_f^2/m_W^2$, and with the remaining form factor contributions being zero. Also, we have defined 
\beq
v=I_3 -2 s_W^2 e_Q\qquad a=I_3 \qquad c^2= v^2 + a^2,
\eeq
with $I_3$ denoting the third component of weak isospin of the fermion and $s_W\equiv \sin\theta_W$ is the Weinberg angle. Notice that we have kept generic the value of the Ricci coupling $\chi$, obtained from the term of improvement for the Higgs doublet. Using the expressions of the form factors given above, in the massless fermion limit, at high momentum transfers $-q^2\gg m_W^2, m_Z^2$, the diagonal cross section is then given by 
\bea
\frac{d\sigma}{d\Omega} &=&\frac{(G M)^2}{\sin^4\frac{\theta}{2}}\frac{G_F}{\sqrt{2}\pi^2}\times \nn\\
&& 2\cos ^2\left(\frac{\theta
   }{2}\right)\left[ -\frac{2}{3}\left( m_W^2(15 + 2 \pi^2) + 15 m_Z^2 (a^2 + v^2)\right) 
- 6 m_W^2 \log\left(-\frac{m_W^2}{q^2}\right)\right. \nn\\
&&\left. - 2 m_w^2\log^2\left(-\frac{m_W^2}{q^2}\right)
-6 m_Z^2 (a^2 + v^2) \log\left(-\frac{m_Z^2}{q^2}\right) 
-4 m_Z^2(a^2 + v^2)  \log^2\left(-\frac{m_Z^2}{q^2}\right)\right].\nn \\
\eea
In this specific limit the result does not depend on the parameter of improvement of the EMT, $\chi$.

\section{Conclusions and perspectives} 
We have presented the structure of the radiative corrections of a scattering of a fermion in an external gravitational field in the Standard Model, extending the original analysis of Berends and Gastmans \cite{Berends:1975ah} which covered the QED case. 
We have used mostly a potential approch, where the gravitational source is assumed to be heavy, such as for a large astrophysical entity, like a star, a planet or a galaxy. In these cases the external field is treated as generating an external Schwartzschild background. This limit is met in a huge variety of physical scatterings, such as those of neutrinos and dark matter particles, but also of leptons.
Our results are rather general and may cover, in principle, a wide variety of cases, having included in our analysis also the corrections coming from a finite geometrical form factor of the external source. The size of these corrections is obviously quite small, but the angles of deflections which underline their expressions, as shown in \cite{Berends:1975ah}, becomes sizeable for ultra high energy cosmic rays, for collisions grazing black holes. In particular, the quadratic growth (with energy) of the angle of deflection of a fermion - discussed in \cite{Berends:1975ah} in the QED case - is therefore expected to contribute significantly to gravitational lensing, also in the weak case. 
These features are of potential interest in the astrophysical context, especially for cosmic rays, which are characterized by huge energies, up to the Greisen-Zatsepin-Kuzmin (GZK) cutoff \cite{Zatsepin:1966jv,Greisen:1966jv}. \\
\\

\centerline{\bf Acknowledgements} 
We thank Mirko Serino for discussions.

\appendix

\section{Appendix. The Schwarzschild background for the scattering of neutrinos and dark matter fermions} 
\label{Schw}
 In this appendix we present a rigorous derivation of the potential given in Eq.~(\ref{pot}).
In the case of a spherically symmetric and stationary source, the ansatz for the metric in polar coordinates, which  takes the form 
\beq
\label{SCH1}
ds^2=e^{\nu(r)} d t^2 - e^{\lambda(r)} dr^2 - r^2 d\theta - r^2\sin^2\theta d\phi^2 
\eeq
introduces two undeterminate functions which depend on a radial coordinate $r$, $\nu(r)$ and $\lambda(r)$.
The Einstein equations in the vacuum determine these two functions in terms of a single parameter $C$ in the form 
\beq
e^{\lambda(r)}=\frac{1}{1 -\frac{C}{r}}, \qquad e^{\nu(r)}=1-\frac{C}{r}
\label{SCH2}
\eeq
as
\beq
\label{ds0}
ds^2=(1-\frac{C}{r}) dt^2 - \frac{1}{1 -\frac{C}{r}} dr^2 - r^2 d\theta - r^2\sin^2\theta d\phi^2 
\eeq

This form needs some manipulation in order to be confronted with the retarded solution generated by $T^{ext}_{\mu\nu}$, characterized by
\bea
g_{00}&=&\eta_{00} +\kappa h_{00}=\left(1- \frac{2 G M}{|\vec{x}|}\right) \nn \\
g_{ii}&=& \eta_{ii}+\kappa h_{ii}=-\left(1 + \frac{2 G M}{|\vec{x}|}\right),
\eea
with the full metric written as
\beq
ds^2\approx\left(1- \frac{2 G M}{|\vec{x}|}\right)dt^2 -\left(1 + \frac{2 G M}{|\vec{x}|}\right)d\vec{x}\cdot d\vec{x}.
\label{SCH3}
\eeq
In order to perform this comparison we need to perform a change of coordinates 
\beq
r=r'\left(1 + \frac{C}{4 r'}\right)^2 
\eeq
which allows to rewrite the vacuum solution (\ref{SCH1}) in the form 
\beq
ds^2=\left( \frac{1- \frac{C}{4 r'}}{1 + \frac{C}{4 r'}}\right)^2 dt^2 - \left( 
1 + \frac{C}{4 r'}\right)^4 (dr'^2 + r'^2 d\theta^2 + r'^2 \sin\theta^2 d\phi^2),
\label{SCH4}
\eeq
which is characterized by a single factor multiplying the spatial part $(d\vec{x}\cdot d\vec{x})$, as in the retarded solution, given in (\ref{SCH3}). Taking the limit $r'\to \infty$, which corresponds to the weak field approximation 
(i.e $b_N\equiv GM/(R c^2) \ll 1$), (\ref{SCH4}) becomes 

\beq
ds^2\approx\left(1- \frac{C}{r'}\right) -\left(1 + \frac{C}{r'}\right) ( dr'^2 + r'^2 d\theta^2 + r'^2 \sin\theta^2 d\phi^2)
\label{SCH5}
\eeq
which allows us to identify $C=2 G M$. Reintroducing $c$, the speed of light, the metric generated by 
(\ref{SCH3}) and (\ref{SCH5}) in the vacuum takes the usual post-newtonian form 
\beq
ds^2\approx \left(1+ 2\frac{\Phi^{ext}}{c^2}\right)dt ^2 -\left(1 - 2\frac{\Phi^{ext}}{c^2}\right) d\vec{x}\cdot d\vec{x}
\label{post}
\eeq
in terms of the Newtonian potential $ \Phi^{ext}(x)=-G M/|\vec{x}|$. This gives 
\beq
h_{00}=h_{ii}=\frac{2 \Phi^{ext}}{c^2 \kappa}
\eeq
or, equivalently 
\beq
h_{\mu\nu}=-\frac{2 \Phi^{ext}}{c^2 \kappa} \bar{S}_{\mu\nu}.
\eeq
\subsection{Interior metric in the weak field limit and the matching} 
The derivation of the form of the metric in the inner region of the distributed source ($r< R$)  requires a further 
examination. We recall that in this case, assuming that the source behaves as a perfect and stationary fluid, the EMT   
takes the form 
\beq
T^{\mu\nu}=(\rho + p)u^\mu u^\nu - p g^{\mu\nu}=diag(\rho,-p,-p,-p)
\eeq
with $u^\mu=(1,\vec{0})$. Setting in Eq.~(\ref{SCH1}) $e^{-\lambda(r)}=1-2\frac{m(r)}{r}$, Einstein's equations combined with the conservation of the energy momentum tensor in the source core take the form 
\bea
\frac{d m(r)}{d r}&=&4 \pi r^2 \rho(r) \nn\\
-\frac{d p(r)}{dr}&=&(\rho(r) +p(r))\left(\frac{m(r)+4 \pi r^3 p(r)}{r^2 (1 - 2{m(r)/r}} \right)\nn \\
\frac{1}{2}\frac{d\nu(r)}{dr}&=&-\frac{1}{\rho(r)+p(r)}\frac{d p(r)}{d r}
\eea
(in geometrized units $G=c=1$). Factors of $G$ and $c$ can be reintroduced by the replacements 
\beq
m\to \frac{G m}{c^2},\qquad \rho\to \frac{G\rho}{c^4},\quad p\to \frac{G p}{c^4}.
\eeq
At this point, if we assume that the density of the fluid be dominated by its rest mass density $(\rho(r)=\mu(r) c^2)$
and we take the nonrelativistic limit by expanding the previous equations in $1/c$, we obtain
\bea
\frac{d m(r)}{d r}&=&4 \pi r^2 \mu(r) \\
-\frac{d p(r)}{dr}&=&\mu(r)\frac{G m(r)}{r^2} \\
\frac{1}{2}\frac{d\nu(r)}{dr}&=&-\frac{1}{\mu(r)}\frac{d p(r)}{d r}
\label{ecco}
\eea
Notice that we have 3 equations and 4 variables. In our case we assume that the mass density is a constant 
$\mu (r)=\mu$ and solve for $m(r), p(r)$ and $\nu(r)$. In the weak field limit $( b_N \ll 1)$  we set $\nu(r)\approx 2 \Phi/c^2$, and Eq.~(\ref{ecco}) becomes 

\beq
\frac{d\Phi^{int}}{d r}=  \frac{G m(r)}{r^2}.
\eeq
The equations can be immediately solved for a body of constant density in the spherical approximation $\mu=M/(4/3 \pi R^3)$ as 
\bea
m(r)&=& \frac{4}{3}\pi \mu r^3  \nn\\
p(r)&=&=-\frac{2}{3}\pi G \mu^2(r^2-R^2) \nn\\
\Phi^{int}(r)&=& -\frac{GM}{2}(\frac{3}{R}- \frac{r^2}{R^3}).
\eea
Introducing $\sigma\equiv b_N/R^2$, the metric takes the form 

\beq
ds^2=\left(1 -3 b_N + \sigma r^2)\right)dt^2 - \frac{1}{\left( 1 - 2 \sigma r^2\right)}dr^2 - r^2 d \theta^2 -
r^2\sin\theta^2 d\phi^2 
\eeq
on which we need to perform a a coordinate redefinition in order to factorize the euclidean differential length 
$d\vec{x}\cdot d\vec{x}$, as in (\ref{SCH4}). To identify this change of coordinates we start considering a rotationally symmetric metric akin (\ref{SCH1}) written, for convenience, in the form 
\beq
ds^2= A(r) dt^2 - B(r)dr^2 -r^2d\theta^2 - r^2\sin^2\theta d\phi^2
\eeq
that we intend to rewrite as 
\beq
ds^2=\tilde{A}(r') dt^2 - \tilde{B}(r')\left(dr'^2 - r'^2 d\theta^2-r'^2\sin^2\theta d\phi^2\right)
\eeq
by a radial redefinition $r'=r'(r)$. We impose the conditions 
\bea
\sqrt{B(r)} dr &=&\sqrt{\tilde{B}(r')} dr' \nn\\
r &=&\sqrt{\tilde{B}(r')} r'
\label{twos}
\eea
with $\tilde{B}(r')$ an unknown function of the new radial coordinate $r'$. In our case 
$A(r)=(1-3 b +\sigma r^2)$ and ${B}(r)=1/(1- 2\sigma r^2)$. The two conditions in (\ref{twos}) generate a separable differential equation which can be integrated as
\beq
\label{solution}
\int_{\tau}^{r'}\frac{d\bf{r'}}{\bf{r'}}=\int_{r_0}^r \frac{d{\bf r}}{{\bf r}(\sqrt{1-2 \sigma{\bf r}^2})}
\eeq
where $\tau=r'(r_0)$. Notice that in this case we keep both $\tau$ and $r_0$ as external parameters, and fix them by the requirements 
that at the core radius $R$ the interior solution coincides with the Schwarzschild solution in the vacuum.
The solution of (\ref{solution}) is 
\beq
r'(r)=\frac{\tau r}{r_0}\frac{(1+\sqrt{1 - 2 \sigma r_0^2})}{(1+\sqrt{1 - 2 \sigma r^2})}
\eeq
which can be inverted in order to determine $r=r(r',\tau,r_0)$. Notice that from (\ref{twos})
 \bea
 \tilde{B}({r'})&=&\frac{r(r')^2}{r'^2}\nn\\
 &=&
\frac{\left(\alpha  {r_0}
   {r'}^3 \sqrt{1-\frac{2 b_N
   {r_0}^2}{R^2}}-\tau ^3
   {r_0} {r'}
   \sqrt{1-\frac{2 b_N
   {r_0}^2}{R^2}}+\tau 
   {r_0} {r'}^3+\tau ^3
   {r_0}
   {r'}\right)^2}{{r'}^2
   \left(\frac{\tau ^4 b_N
   {r_0}^2}{R^2}+\frac{b_N
   {r_0}^2
   {r'}^4}{R^2}-\frac{2 \tau
   ^2 b_N {r_0}^2
   {r'}^2}{R^2}+2 \tau ^2
   {r'}^2\right)^2}
\eea
which is valid beyond the weak field limit. In the weak field limit 
\beq
\tilde{B}(r')=\frac{{r_0}^2}{\tau
   ^2}+\frac{b_N {r_0}^4
   \left(\tau^2-{r'}^2\right)}{\tau^4
   R^2}+O\left(b_N^2\right)
 \eeq
The expression $\tilde{A}{(r')}$ valid at $O(b_N)$ is 
\beq
\tilde{A}(r')=b \left(\frac{{r_0}^2
   {r'}^2}{\tau ^2
   R^2}-3\right)+1 + O(b_N^2)
 \eeq
At this point we introduce the matching conditions at $O(b_N)$ between the interior and the exterior potentials at 
$r=R$
\bea
\tilde{A}(r'=R)&=& 1 + 2\frac{\Phi^{ext}(R)}{c^2}=1 - 2 b_N\nn\\
 \tilde{B}(r'=R)&=& 1- 2\frac{\Phi^{ext}(R)}{c^2} =1+2 b_N
 \eea
 which fix the parameters as $\tau=r_0=\sqrt{3} R$. At the same time we obtain at the same order
 \beq
 \tilde{A}(r')=1 + 2\frac{\Phi^{int}}{c^2}, \qquad \tilde{B}(r')=1 -2\frac{\Phi^{int}}{c^2}.
 \eeq
 This shows that at $O(b_N)$ the metric can be expressed as in (\ref{pot})  as expected, on the other hand, from Newtonian dynamics. The approach that we have followed can be easily extended to higher order in $b_N$ in the "tt" component $A(r)$, while the expression of $\tilde{B}{(r')}$ remains always valid.\\
Notice that in the computation of the geometrical form factors of the source we need the integral
\beq
\int_{R_>} d^3\vec{x} \frac{e^{i \vec{q}\cdot \vec{x}}}{|\vec{x}|}\theta(|\vec{x}|-R)
\eeq
which requires a regularization at large momentum, due to the presence of an oscillating factor. The simplest way avoid any regularization is to redefine it as difference between the contribution of the entire region and that of the interior region ($r < R$), the latter being given by 
\beq
\int_{R_<} d^3\vec{x} \frac{e^{i \vec{q}\cdot \vec{x}}}{|\vec{x}|}\theta(R- |\vec{x}|)=\frac{4 \pi}{\vec{q}^2}\left( 
1- \cos (|\vec{q}| R)\right).
\eeq
Therefore we obtain
\bea
\int_{R_>} d^3\vec{x} \frac{e^{i \vec{q}\cdot \vec{x}}}{|\vec{x}|}\theta(|\vec{x}|-R) &=&\int d^3\vec{x} \frac{e^{i \vec{q}\cdot \vec{x}}}{|\vec{x}|}
- \int_{R_<} d^3\vec{x} \frac{e^{i \vec{q}\cdot \vec{x}}}{|\vec{x}|}\theta(R- |\vec{x}|) \nn \\
&=& \frac{4 \pi}{\vec{q}^2} - \frac{4 \pi}{\vec{q}^2}\left(1- \cos (|\vec{q}| R)\right)\nn \\
&=&\frac{4 \pi}{\vec{q}^2}\cos (|\vec{q}| R).
\eea

\section{ List of $g$ functions and $c$ coefficients}
The $c$ coefficients of the flavour-changing cross section Eq.~(\ref{non1}), in the case of equal mass fermions, are given by 

\begin{table}[here]
\label{cg}
\begin{tabular}{llll}
$c_1=72 m^2-18 q^2$ & $c_3=32 m^2-8 q^2$ & $c_4=\frac{1}{2 }(4 m^2-q^2)^3$ & $c_5=\frac{1}{2} (4 m^2-q^2) q^2$   \\
$c_7=8 q^2$& $c_8=\frac{1}{2} q^2 (-4 m^2+q^2)^2$ & $c_9= \frac{1}{2}(q^2)^3$ & 
$c_{11}=\frac{2 (m^2+2 m_w^2)^2 (4 m^2-q^2)}{m_w^4}$  \\
$d_{1\,\, 3}=96 m^2- 24 q^2$ & $d_{1\,4}=6 (-4 m^2+q^2)^2$ & $d_{1\,5}= 6 (4 m^2-q^2) q^2$ &
$d_{1 \,\,11}=\frac{12 (m^2+2 m_w^2) (4 m^2-q^2)}{m_w^2}$ \\
$d_{3\,\, 4}=4 (-4 m^2+q^2)^2$ & $d_{3\, \,5}= q^2 d_{3\, 4}$ & $d_{3 \,\,11}=\frac{2}{3}d_{1 \,11}$ & $d_{4\,\, 5}=q^2 (-4 m^2+q^2)^2$   \\
$d_{4 \,\, 11}=\frac{1}{6} d_{1 \,\, 11}$ & $d_{5\, 11}=\frac{q^2}{4} d_{3\, \,11}$ & $d_{7\,\, 8}=4 (4 m^2-q^2) q^2$ 
& $d_{7\,\, 9}= 4 (q^2)^2$  \\
$d_{8\, \,9}=(q^2)^2 ( 4 m^2 - q^2)$.
\end{tabular}
\end{table}
\beq
\eeq

The functions $g_i(x)$  are given by
\bea
g_a(x)&=& 
\frac{1}{36 \left(x-1\right)^4}
\left[44-194x+243x^2-98x^3 +5x^4
\right.
\nonumber\\
&+&\left. 6x\left(2-15x+10 x^2\right)
\log{(x)}\right]
\nonumber 
\eea
\bea
g_b(x)&=& 
\frac{1}{6 \left(x-1\right)^4}
\left[8-14x+21x^2-14x^3-x^4
+2x\left(4+3x+2 x^3\right)
\log{(x)}\right]
\nonumber 
\eea
\bea
g_c(x)&=& 
\frac{1}{36 \left(x-1\right)^6}
\left[2+9x-152x^2+88x^3+54x^4-x^5
\right.
\nonumber\\
&-&\left. 12x\left(-1+5x+10x^2+x^3\right)
\log{(x)}\right]
\nonumber\\
g_{d}(x)&=& 
\frac{1}{72 \left(x-1\right)^6}
\left[6-83x+200x^2+12x^3-142x^4 +7x^5 
\right.
\nonumber\\
&-& \left. 12x\left(1+4x-18 x^2-2x^3\right)
\log{(x)}\right]
\nonumber \\
g_g(x)&=& 
\frac{1}{360\left(x-1\right)^6}
\left[
106+245x-240x^2+20x^3+70x^4-201x^5
\right.
\nonumber\\
&+& \left. 12\left(2+23x+18x^2+8x^3+18x^4+6x^5\right)
\log{(x)}\right]
\nonumber \\
g_{h}(x)&=& 
\frac{1}{240\left(x-1\right)^6}
\left[5\left(
-6+41x+64x^2-180x^3+70x^4+11x^5\right)
\right.
\nonumber\\
&+& \left. 4\left(-6+21x+76x^2+36x^3-84x^4+2x^5\right)
\log{(x)}\right].
\eea


 \end{document}